\documentclass[longauth]{aa}
\usepackage{graphicx}
\usepackage{txfonts}
\usepackage{xcolor}
\usepackage{hyperref}
\hypersetup{colorlinks, linkcolor=blue, citecolor=blue, urlcolor=blue}
\newcommand{\te}{t_{\rm E}}
\newcommand{\thetae}{\theta_{\rm E}}

\newcommand{\pie}{\pi_{\rm E}}

\newcommand{\dl}{D_{\rm L}}
\newcommand{\ds}{D_{\rm S}}

\definecolor{brown}{rgb}{0.59, 0.29, 0.0}
\definecolor{darkgreen}{rgb}{0.0, 0.42, 0.24}
\definecolor{darkblue}{rgb}{0.01, 0.31, 0.59}
\definecolor{darkblue}{rgb}{0.0, 0.25, 0.42}
\definecolor{blue}{rgb}{0.0,0.0,1.0}
\definecolor{green}{rgb}{0.0,1.0,0.0}

\begin{document}

\title{
Six microlensing planets detected via sub-day signals during the 2023 -- 2024 season
}
\titlerunning{Microlensing planets with sub-day signals}

\author{
% leading author -----------------------------
     Cheongho~Han\inst{\ref{cbnu}}
\and Chung-Uk~Lee\inst{\ref{kasi}\thanks{\tt leecu@kasi.re.kr}} 
\and Andrzej~Udalski\inst{\ref{warsaw}} 
\and Ian~A.~Bond\inst{\ref{massey}}
\\
(Leading authors)
\\
% KMTNet ---------------------------
     Michael~D.~Albrow\inst{\ref{canterbury}}   
\and Sun-Ju~Chung\inst{\ref{kasi}}      
\and Andrew~Gould\inst{\ref{osu}}      
\and Youn~Kil~Jung\inst{\ref{kasi},\ref{ust}} 
\and Kyu-Ha~Hwang\inst{\ref{kasi}} 
\and Yoon-Hyun~Ryu\inst{\ref{kasi}} 
\and Yossi~Shvartzvald\inst{\ref{weizmann}}   
\and In-Gu~Shin\inst{\ref{cfa}} 
\and Jennifer~C.~Yee\inst{\ref{cfa}}   
\and Weicheng~Zang\inst{\ref{cfa},\ref{tsinghua}}     
\and Hongjing~Yang\inst{\ref{tsinghua}}     
\and Sang-Mok~Cha\inst{\ref{kasi},\ref{kyunghee}} 
\and Doeon~Kim\inst{\ref{cbnu}}
\and Dong-Jin~Kim\inst{\ref{kasi}} 
\and Seung-Lee~Kim\inst{\ref{kasi}} 
\and Dong-Joo~Lee\inst{\ref{kasi}} 
\and Yongseok~Lee\inst{\ref{kasi},\ref{kyunghee}} 
\and Byeong-Gon~Park\inst{\ref{kasi}} 
\and Richard~W.~Pogge\inst{\ref{osu}}
\\
(The KMTNet Collaboration)
\\
%%%% OGLE ---------------------------
     Przemek~Mr{\'o}z\inst{\ref{warsaw}} 
\and Micha{\l}~K.~Szyma{\'n}ski\inst{\ref{warsaw}}
\and Jan~Skowron\inst{\ref{warsaw}}
\and Rados{\l}aw~Poleski\inst{\ref{warsaw}} 
\and Igor~Soszy{\'n}ski\inst{\ref{warsaw}}
\and Pawe{\l}~Pietrukowicz\inst{\ref{warsaw}}
\and Szymon~Koz{\l}owski\inst{\ref{warsaw}} 
\and Krzysztof~A.~Rybicki\inst{\ref{warsaw},\ref{weizmann}}
\and Patryk~Iwanek\inst{\ref{warsaw}}
\and Krzysztof~Ulaczyk\inst{\ref{warwick}}
\and Marcin~Wrona\inst{\ref{warsaw},\ref{villanova}}
\and Mariusz~Gromadzki\inst{\ref{warsaw}}          
\and Mateusz~J.~Mr{\'o}z\inst{\ref{warsaw}} 
\and Micha{\l} Jaroszy{\'n}ski\inst{\ref{warsaw}}
\and Marcin Kiraga\inst{\ref{warsaw}}
\\
(The OGLE Collaboration)
\\
%%%% MOA ---------------------------
     Fumio~Abe\inst{\ref{nagoya}}
\and David~P.~Bennett\inst{\ref{nasa},\ref{maryland}}
\and Aparna~Bhattacharya\inst{\ref{nasa},\ref{maryland}}
\and Akihiko~Fukui\inst{\ref{tokyo-earth},}\inst{\ref{spain}}
\and Ryusei~Hamada\inst{\ref{osaka}}
\and Stela~Ishitani~Silva\inst{\ref{nasa},\ref{catholic}}  
\and Yuki~Hirao\inst{\ref{tokyo-ast}}
\and Naoki~Koshimoto\inst{\ref{tokyo-ast}}
\and Yutaka~Matsubara\inst{\ref{nagoya}}
\and Shota~Miyazaki\inst{\ref{osaka}}
\and Yasushi~Muraki\inst{\ref{nagoya}}
\and Tutumi~Nagai\inst{\ref{osaka}}
\and Kansuke~Nunota\inst{\ref{osaka}}
\and Greg~Olmschenk\inst{\ref{nagoya}}
\and Cl{\'e}ment~Ranc\inst{\ref{sorbonne}}
\and Nicholas~J.~Rattenbury\inst{\ref{auckland}}
\and Yuki~Satoh\inst{\ref{osaka}}
\and Takahiro~Sumi\inst{\ref{osaka}}
\and Daisuke~Suzuki\inst{\ref{osaka}}
\and Sean K. Terry\inst{\ref{nasa}, \ref{maryland}}
\and Paul~J.~Tristram\inst{\ref{john}}
\and Aikaterini~Vandorou\inst{\ref{nasa},\ref{maryland}}
\and Hibiki~Yama\inst{\ref{osaka}}
}

\institute{
      Department of Physics, Chungbuk National University, Cheongju 28644, Republic of Korea                                                          \label{cbnu}     
\and  Korea Astronomy and Space Science Institute, Daejon 34055, Republic of Korea                                                                    \label{kasi}   
\and  Astronomical Observatory, University of Warsaw, Al.~Ujazdowskie 4, 00-478 Warszawa, Poland                                                      \label{warsaw}   
\and  Institute of Natural and Mathematical Science, Massey University, Auckland 0745, New Zealand                                                    \label{massey}    
\and  University of Canterbury, Department of Physics and Astronomy, Private Bag 4800, Christchurch 8020, New Zealand                                 \label{canterbury}  
\and  Department of Astronomy, Ohio State University, 140 West 18th Ave., Columbus, OH 43210, USA                                                     \label{osu} 
\and  University of Science and Technology, Daejeon 34113, Republic of Korea                                                                          \label{ust}
\and  Department of Particle Physics and Astrophysics, Weizmann Institute of Science, Rehovot 76100, Israel                                           \label{weizmann}   
\and  Center for Astrophysics $|$ Harvard \& Smithsonian 60 Garden St., Cambridge, MA 02138, USA                                                      \label{cfa}  
\and  Department of Astronomy and Tsinghua Centre for Astrophysics, Tsinghua University, Beijing 100084, China                                        \label{tsinghua} 
\and  School of Space Research, Kyung Hee University, Yongin, Kyeonggi 17104, Republic of Korea                                                       \label{kyunghee}     
\and  Department of Physics, University of Warwick, Gibbet Hill Road, Coventry, CV4 7AL, UK                                                           \label{warwick}
\and  Villanova University, Department of Astrophysics and Planetary Sciences, 800 Lancaster Ave., Villanova, PA 19085, USA                           \label{villanova} 
\and  Institute for Space-Earth Environmental Research, Nagoya University, Nagoya 464-8601, Japan                                                     \label{nagoya}     
\and  Code 667, NASA Goddard Space Flight Center, Greenbelt, MD 20771, USA                                                                            \label{nasa} 
\and  Department of Astronomy, University of Maryland, College Park, MD 20742, USA                                                                    \label{maryland}  
\and  Department of Earth and Planetary Science, Graduate School of Science, The University of Tokyo, 7-3-1 Hongo, Bunkyo-ku, Tokyo 113-0033, Japan   \label{tokyo-earth} 
\and  Instituto de Astrof{\'i}sica de Canarias, V{\'i}a L{\'a}ctea s/n, E-38205 La Laguna, Tenerife, Spain                                            \label{spain} 
\and  Department of Earth and Space Science, Graduate School of Science, Osaka University, Toyonaka, Osaka 560-0043, Japan                            \label{osaka}  
\and  Department of Physics, The Catholic University of America, Washington, DC 20064, USA                                                            \label{catholic} 
\and  Institute of Astronomy, Graduate School of Science, The University of Tokyo, 2-21-1 Osawa, Mitaka, Tokyo 181-0015, Japan                        \label{tokyo-ast}
\and  Sorbonne Universit\'e, CNRS, UMR 7095, Institut d'Astrophysique de Paris, 98 bis bd Arago, 75014 Paris, France                                  \label{sorbonne}
\and  Department of Physics, University of Auckland, Private Bag 92019, Auckland, New Zealand                                                         \label{auckland}    
\and  University of Canterbury Mt.~John Observatory, P.O. Box 56, Lake Tekapo 8770, New Zealand                                                       \label{john}  
}                                                                                                                                                       
\date{Received ; accepted}

% \abstract{}{}{}{}{} 
% 5 {} token are mandatory
\abstract
{}
{
We present analyses of six microlensing events: KMT-2023-BLG-0548, KMT-2023-BLG-0830, 
KMT-2023-BLG-0949, KMT-2024-BLG-1281, KMT-2024-BLG-2059, and KMT-2024-BLG-2242. 
These were identified in KMTNet data from the 2023 -- 2024 seasons, selected for 
exhibiting anomalies shorter than one day -- potential signatures of low-mass 
planetary companions. Motivated by this, we conducted detailed investigations to 
characterize the nature of the observed perturbations.
}
{
Detailed modeling of the light curves reveals that the anomalies in all six 
events are caused by planetary companions to the lenses. The brief durations of 
the anomalies are attributed to various factors: a low planet-to-host mass ratio 
(KMT-2024-BLG-2059, KMT-2024-BLG-2242), a wide planet-host separation 
(KMT-2023-BLG-0548), small and elongated caustics restricting the source's 
interaction region (KMT-2023-BLG-0830, KMT-2024-BLG-1281), and a partial caustic 
crossing (KMT-2023-BLG-0949).
}
{
We estimated the physical parameters of the lens systems using Bayesian analysis. 
For KMT-2023-BLG-0548, the posterior distribution of the lens mass shows two 
distinct peaks: a low-mass solution indicating a sub-Jovian planet orbiting an M 
dwarf in the Galactic disk, and a high-mass solution suggesting a super-Jovian planet 
around a K-type dwarf in the bulge.  KMT-2023-BLG-0830 hosts a Neptune-mass planet 
orbiting an M dwarf in the Galactic bulge.  KMT-2023-BLG-0949 involves a super-Jovian 
planet orbiting a $\sim 0.5~M_\odot$ host located at $\sim 6$ kpc. KMT-2024-BLG-2059Lb 
is a super-Earth with a mass about seven times that of Earth, orbiting an early M 
dwarf of $\sim 0.5~M_\odot$.  KMT-2024-BLG-1281L hosts a planet slightly more massive 
than Neptune, orbiting an M dwarf of $\sim0.3~M_\odot$. The short timescale and small 
angular Einstein radius of KMT-2024-BLG-2242 suggest a $\sim 0.07~M_\odot$ primary, 
likely a brown dwarf, with a Uranus/Neptune-mass planet.
}
{}

\keywords{planets and satellites: detection -- gravitational lensing: micro}

\maketitle

\section{Introduction} \label{sec:one}

Planetary microlensing signals appear as brief, discontinuous anomalies in the 
smooth, symmetric lensing light curve of the planet-hosting star \citep{Mao1991, 
Gould1992}.  These signals arise when the source of a lensing event passes close 
to the caustic induced by the planet. Caustics are regions in the source plane 
where the magnification of a point source becomes infinite. When a lens system 
consists of a planet and its host star, two distinct types of caustics form. 
The first is a small central caustic located near the primary lens, while the 
second is a peripheral caustic situated at a distance of ${\bf s} - 1/{\bf s}$ 
from the host star. Due to their respective locations, the central caustic 
produces anomalies near the peak of a high-magnification event light curve, 
whereas the peripheral caustic primarily generates deviations in the outer 
regions of the light curve.

The duration of a planetary microlensing signal is primarily determined by the 
size of the caustic. For peripheral caustics, the caustic size scales with the 
square root of the planet-to-host mass ratio \citep{Han2006}, whereas for central 
caustics, it scales linearly with the mass ratio \citep{Chung2005}. As a result, 
lower-mass planets produce smaller caustics, leading to shorter-duration planetary 
signals. Therefore, brief planetary anomalies in microlensing light curves are 
more likely to originate from low-mass planets.

However, short signals in lensing light curves can arise from factors other 
than the presence of a low-mass planet.  For planet-induced caustics, their 
size depends not only on the mass ratio but also on the projected separation 
between the planet and its host star.  As a result, if this separation deviates 
significantly from the Einstein radius of the lens, a small caustic may form.  
Additionally, the signal duration can be shortened if the caustic is highly 
elongated and the source crosses it along a direction with a small cross-sectional 
area. A similar effect occurs when the source passes near the edge of the caustic, 
leading to a reduced signal duration.  Moreover, short-term anomalies can also 
originate from non-planetary causes. One such case involves a binary lens system 
in which the separation between the two components is either much smaller or much 
larger than the Einstein radius \citep{Han2008}. It is also well known that a 
faint companion to the source star can produce a short-term anomaly in the 
lensing light curve \citep{Gaudi1998, Gaudi2004}.

In this paper, we analyze six microlensing events: KMT-2023-BLG-0548, 
KMT-2023-BLG-0830, KMT-2023-BLG-0949, KMT-2024-BLG-1281, KMT-2024-BLG-2059, 
and KMT-2024-BLG-2242.  These events share a common characteristic of displaying 
anomalies in their light curves that last less than a day.  Since such short-term 
anomalies in a lensing light curve suggest the possible presence of a low-mass 
planetary companion to the lens, we conducted detailed analyses of these events 
to investigate the origin and nature of the anomalies.

For the presentation of the analyses, the paper is structured as follows. In 
Sect.~\ref{sec:two}, we outline the process of identifying the anomalies in the 
lensing light curves of the events analyzed in this study, along with the data 
employed for the analysis. In Sect.~\ref{sec:three}, we provide an overview of 
the fundamentals of planetary microlensing and detail the modeling procedure 
employed to interpret the anomalies observed in the lensing light curves. Within 
the six subsections of Sect.~\ref{sec:four}, we present the modeling results 
for each individual event and discuss the origins of the anomalies. In Sect. 
\ref{sec:five}, we characterize the source stars for each event and determine 
the angular Einstein radii for the events for which finite-source effects were 
detected. In Sect.~\ref{sec:six}, we estimate the masses and distances of the 
lens systems by performing Bayesian analyses based on the observables derived 
from the analyses. Finally, in Sect.~\ref{sec:seven}, we summarize the findings 
and present our conclusions.

\section{Observations and data} \label{sec:two}

The lensing events analyzed in this study were identified by examining 
microlensing events detected by the Korea Microlensing Telescope Network 
\citep[KMTNet;][]{Kim2016} during the 2023 and 2024 seasons.  The goal was to 
identify lensing events that displayed very short-term anomalies in their light 
curves.  From this examination, we found six lensing events, KMT-2023-BLG-0548, 
KMT-2023-BLG-0830, KMT-2023-BLG-0949, KMT-2024-BLG-1281, KMT-2024-BLG-2059, and 
KMT-2024-BLG-2242, that each displayed short-term anomalies lasting less than a 
day.  These events were found with the application of the KMTNet EventFinder system 
\citep{Kim2018} applied to the KMTNet data acquired during the corresponding seasons.  
The anomalies in these events were identified with the use of the semi-automatic 
AnomalyFinder algorithm \citep{Zang2021}.

We examined the availability of additional data from other microlensing surveys. 
Among the events identified by the KMTNet survey, KMT-2023-BLG-0949 and 
KMT-2024-BLG-2059 were also observed by the Optical Gravitational Lensing Experiment 
\citep[OGLE;][]{Udalski2015}, while KMT-2024-BLG-1281 was observed by the Microlensing 
Observations in Astrophysics \citep[MOA;][]{Bond2001, Sumi2003}. KMT-2024-BLG-2242 
was covered by both the OGLE and MOA surveys. For these events, we carried out the 
analysis using the combined data sets from the respective surveys.

The data for the events were acquired using wide-field telescopes of the survey 
groups specifically designed for microlensing surveys. The KMTNet group operates 
three identical telescopes strategically positioned across the Southern Hemisphere 
to facilitate continuous observations. These telescopes are located at the Siding 
Spring Observatory in Australia (KMTA), the Cerro Tololo Inter-American Observatory 
in Chile (KMTC), and the South African Astronomical Observatory in South Africa (KMTS). 
Each KMTNet telescope features a 1.6-meter aperture and is equipped with a camera 
that provides a field of view of 4 square degrees. The OGLE telescope, situated at 
the Las Campanas Observatory in Chile, has a 1.3-meter aperture and a camera offering 
a field of view of 1.4 square degrees.  The MOA survey conducts observations using 
a 1.8-meter telescope stationed at Mt. John University Observatory in New Zealand, 
featuring a wide-field camera capable of imaging an area of 2.2 square degrees in 
a single exposure.

The KMTNet and OGLE surveys primarily conducted observations in the $I$ band, 
whereas the MOA survey used a custom MOA-$R$ filter, which covers a wavelength 
range of 609 to 1109 nm.  In all three surveys, a subset of images was also taken 
in the $V$ band to enable the measurement of source colors.  The photometry of the 
events was conducted using the pipelines of the respective groups.  To ensure the 
optimal quality of data, the KMTNet data were reprocessed using the code developed 
by \citet{Yang2024}. The photometric errors estimated by the automated pipeline 
were rescaled to align the errors with the observed scatter in the data and to 
ensure that the $\chi^2$ value of the model matched unity for each dataset.  This 
rescaling procedure was performed in accordance with the methodology outlined in 
\citet{Yee2012}.

% Table 1 ------------------------------------------------
\begin{table*}[t]
%\small
%\centering
\caption{Lensing parameters of solutions for KMT-2023-BLG-0548.  \label{table:one}}
\begin{tabular}{llllllllllcc}
%\begin{tabular}{\columnwidth}{@{\extracolsep{\fill}}lllcc}
\hline\hline
\multicolumn{1}{c}{Parameter}  &
\multicolumn{2}{c}{2L1S (Local A)}   &
\multicolumn{2}{c}{2L1S (Local B) }   &
\multicolumn{1}{c}{1L2S}       \\
\multicolumn{1}{c}{}           &
\multicolumn{1}{c}{Close}      &
\multicolumn{1}{c}{Wide }      &
\multicolumn{1}{c}{Close}      &
\multicolumn{1}{c}{Wide }      &
\multicolumn{1}{c}{}           \\
\hline
 $\chi^2$                  &   $1911.1             $   &  $1910.2            $   &   $1911.6            $  &  $1911.8            $   &   $1922.8            $  \\
 $t_0$ (HJD$^\prime$)      &   $56.3041 \pm 0.0047 $   &  $56.3052 \pm 0.0036$   &   $56.3051 \pm 0.0032$  &  $56.3057 \pm 0.0033$   &   $56.3172 \pm 0.0023$  \\
 $u_0$ ($10^{-3}$)         &   $1.27 \pm 0.39      $   &  $0.98 \pm 0.28     $   &   $1.14 \pm 0.17     $  &  $1.00 \pm 0.15     $   &   $-0.23 \pm 0.78    $  \\
 $\te$ (days)              &   $22.65 \pm 4.7      $   &  $27.05 \pm 3.63    $   &   $24.24 \pm 2.95    $  &  $26.95 \pm 2.68    $   &   $20.70 \pm 2.94    $  \\
 $s$                       &   $0.42 \pm 0.11      $   &  $2.43 \pm 0.43     $   &   $0.39 \pm 0.08     $  &  $2.77 \pm 0.51     $   &     --                  \\
 $q$ ($10^{-3}$)           &   $5.32 \pm 2.40      $   &  $4.51 \pm 2.04     $   &   $3.97 \pm 2.18     $  &  $4.55 \pm 2.12     $   &     --                  \\
 $\alpha$ (rad)            &   $2.28 \pm 0.10      $   &  $2.27 \pm 0.07     $   &   $0.16 \pm 0.05     $  &  $0.17 \pm 0.06     $   &     --                  \\
 $\rho$ ($10^{-3}$)        &   $1.52 \pm 0.45      $   &  $1.37 \pm 0.36     $   &   $1.59 \pm 0.25     $  &  $1.43 \pm 0.22     $   &   $2.45 \pm 0.60     $  \\
 $t_{0,2}$ (HJD$^\prime$)  &    --                     &   --                    &    --                   &   --                    &   $55.89 \pm 2.19    $  \\
 $u_{0,2}$ ($10^{-3}$)     &    --                     &   --                    &    --                   &   --                    &   $0.005 \pm 0.315   $  \\
 $q_F$                     &    --                     &   --                    &    --                   &   --                    &   $0.14 \pm 1.18     $  \\
\hline
\end{tabular}
\tablefoot{ ${\rm HJD}^\prime = {\rm HJD}- 2460000$. }
\end{table*}
% --------------------------------------------------------

\section{Light curve modeling} \label{sec:three}

The modeling process involved determining the lensing parameters that best characterize 
the observed lensing light curves. For a single-lens single-source (1L1S) event, the 
light curve is described by three parameters. The first two parameters, $(t_0, u_0)$, 
represent the time of the closest approach between the lens and source and their 
separation at that moment. The third parameter, $\te$, denotes the event timescale, 
which is defined as the duration required for the source to traverse the Einstein 
radius ($\thetae$) of the lens. The parameter $u_0$ is expressed in units of $\thetae$.

For a binary-lens single-source (2L1S) event, which includes an additional lens 
component compared to the 1L1S case, four additional parameters $(s, q, \alpha, 
\rho)$ are required for modeling.  Here, $s$ and $q$ represent the projected 
separation and mass ratio between the two lens components, respectively. The 
parameter $\alpha$ denotes the angle between the source trajectory and the binary 
lens axis, while $\rho$ is the ratio of the angular source radius ($\theta_*$) to 
the Einstein radius, given by $\rho = \theta_*/\thetae$. The parameter $\rho$ is 
necessary to account for finite-source effects, which arise when the source passes 
very close to or crosses the caustic, a frequent characteristic feature of 2L1S 
events.  The binary separation $s$ is expressed in units of $\thetae$.

Similar to the 2L1S case, modeling a single-lens binary-source (1L2S) event requires 
additional parameters.  These parameters are $(t_{0,2}, u_{0,2}, q_F)$, where the 
first two parameters denote the approach time and separation of the second source 
relative to the lens, respectively, while the last parameter represents the flux 
ratio between the secondary and primary sources \citep{Hwang2013}.  For a summary 
of lensing parameters across different lens-system configurations, refer to Table 
2 of \citet{Han2023}.

The 2L1S modeling was conducted using a hybrid approach that combines grid and downhill
optimization methods. Due to the large number of model parameters, it is challenging to 
identify a lensing solution using a purely grid-based approach.  Therefore, we explored 
the parameter space for $s$ and $q$, for which the lensing magnification changes abruptly 
with variations in these parameters, using a grid approach, while the remaining parameters, 
for which magnification varies smoothly, were optimized using downhill methods with multiple 
starting values of $\alpha$.  Another reason for using the grid approach to search for $s$ 
and $q$ is to identify degenerate solutions that might be  present in the plane of these 
parameters.  To compute finite-source magnifications, we employed the map-making technique 
based on the ray-shooting method \citep{Dong2006}.  In this process, variations in the 
surface brightness profile of the source star due to limb darkening were taken into 
account by considering the spectral type of the star. Details on the determination of 
the source type are provided in Sect.~\ref{sec:five}.

The 1L2S modeling was carried out by first applying 1L1S modeling to the light curve 
while excluding the part around the anomaly.  In the next step, the binary-source 
parameters responsible for the anomaly were determined using a downhill approach, with 
initial parameters set based on the strength and time of the anomaly.

\section{Analyses of individual events} \label{sec:four}

In the following subsections, we provide a detailed analysis of each individual event and 
present the results obtained. For cases with multiple possible interpretations, we discuss 
the sources of the degeneracy.

\subsection{KMT-2023-BLG-0548} \label{sec:four-one}

The lensing event KMT-2023-BLG-0548 occurred on a faint star at equatorial coordinates 
$({\rm RA}, {\rm DEC})_{\rm J2000} = (18$:01:23.43, -27:06:24.30), corresponding to 
Galactic coordinates $(l, b) = (3^\circ\hskip-2pt .3426$, -$2^\circ\hskip-2pt .0710)$. 
The $I$-band extinction toward this field is $A_I = 2.24$, and the baseline source 
magnitude prior to lensing magnification was $I_{\rm base} = 20.01$.

% Figure 1 ------------------------------------------------------
\begin{figure}[t]
\includegraphics[width=\columnwidth]{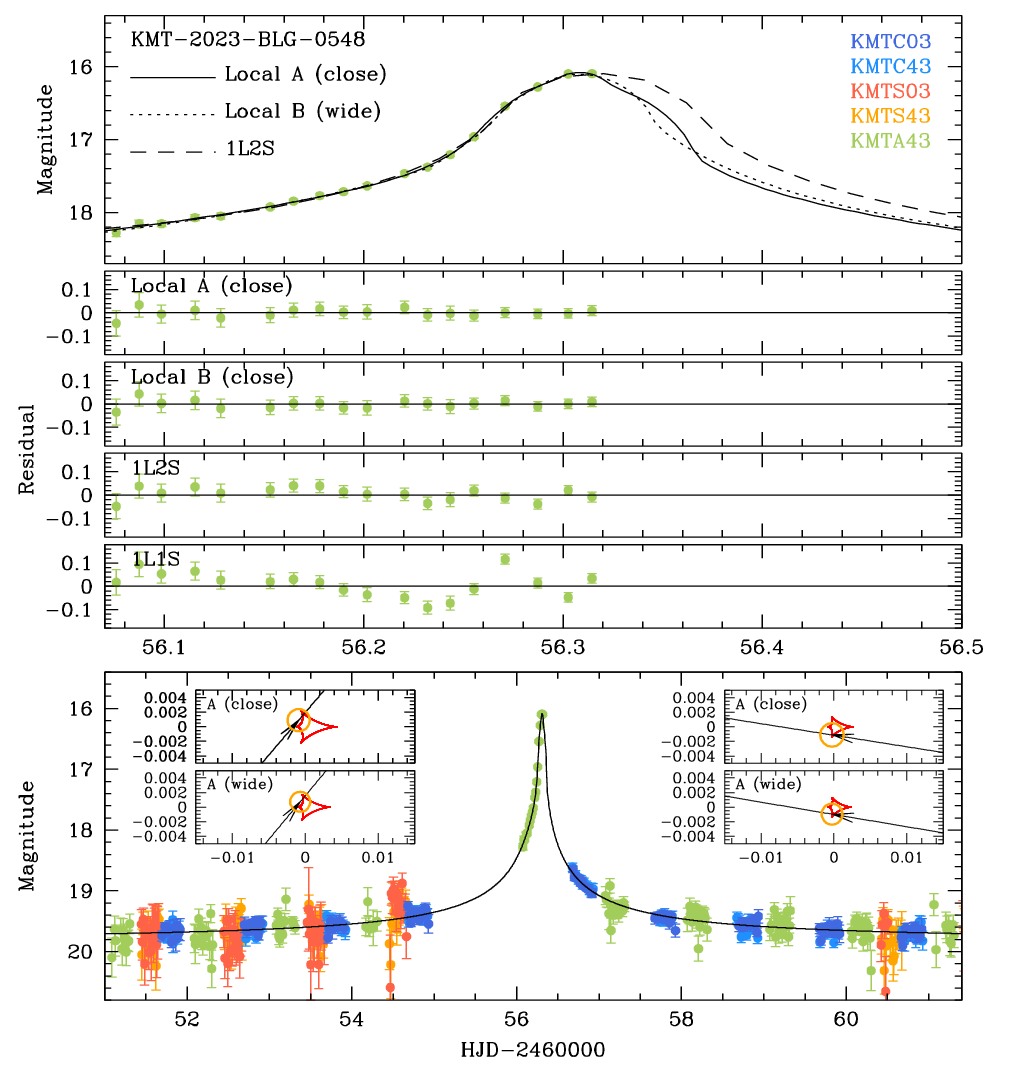}
\caption{
Light Curve of the microlensing event KMT-2023-BLG-0548.  The bottom panel displays the 
full light curve, while the upper panels present a zoomed-in view of the peak region along 
with the residuals for four different models.  The four insets in the bottom panel illustrate 
the lens-system configurations corresponding to the four degenerate 2L1S solutions.  In each 
inset, the closed figure composed of concave curves represents the caustics, and the arrowed 
line indicates the source trajectory.  The orange circle, scaled to represent the angular 
size of the source, on the trajectory marks the source’s position at the moment of peak 
anomaly.
}
\label{fig:one}
\end{figure}
% --------------------------------------------------------------

The KMTNet survey first detected the lensing-induced flux magnification of the source on 
2023 April 24 (corresponding to the abridged Heliocentric Julian Date ${\rm HJD}^\prime 
\equiv {\rm HJD} - 2460000 = 58$), two days after the event had reached its very high peak 
magnification of $A_{\rm max} \sim 940$. The source lies in the overlapping region of the 
KMTNet prime fields BLG03 and BLG43, which were observed with a cadence of 0.5 hours per 
field, corresponding to a combined cadence of 0.25 hours.

Figure~\ref{fig:one} shows the lensing light curve of the event, which appears 
consistent with a high-magnification 1L1S event. However, model fitting to the 1L1S framework 
revealed subtle deviations near the peak of the light curve. These anomalies were brief, 
lasting only several hours, and were captured by KMTA observations from the BLG43 field. 
Although the BLG03 field also covered the peak region, its data were excluded from the 
analysis due to large photometric uncertainties.

Since a central anomaly in a lensing light curve can be attributed either to a planet orbiting 
the lens \citep{Griest1998} or to a binary companion located either close to or far from the 
lens \citep{Han2008}, we conducted 2L1S modeling of the light curve. This analysis resulted in 
four sets of planetary solutions, with the full lensing parameters provided in Table~\ref{table:one}. 
The first two solutions share a similar source trajectory angle of $\alpha \sim 2.27$ radians 
(local A), while the other two have an angle of $\alpha \sim 0.16$ radians (local B), suggesting 
that the degeneracy between solutions in the local A and B groups is coincidental. In contrast, 
the planetary separations within each local group follow the relation $s_{\rm close} \sim 
1/s_{\rm wide}$, implying that the similarity between the model light curves of the degenerate 
solutions is a result of the close-wide degeneracy \citep{Dominik1999}.  Here, the terms 
``close'' and ``wide'' refer to solutions with $s < 1$ and $s > 1$, respectively.  Regardless 
of the specific solution, the estimated event timescale is $\sim 25$ days, and the planet-to-host 
mass ratios are on the order of $10^{-3}$, indicating that the companion to the lens is likely 
a planetary object.

In the top panels of Figure~\ref{fig:one}, we present the model curves for the solutions in 
the region surrounding the anomaly. For each pair of local A and B solutions, we display the 
model curve of the close solution as the representative, since the model light curves from 
the close and wide solutions are nearly identical.  In contrast, the model curves for the 
local A and B solutions start to diverge beyond ${\rm HJD}^\prime \geq 53.32$, a time interval 
during which no data were available to constrain the anomaly.  This divergence provides further 
confirmation that the degeneracy between the two solutions is accidental, and the ambiguity 
could have been resolved if the observations had extended to this region, which would have 
allowed for a clearer distinction between the solutions.

% Table 2 ------------------------------------------------
\begin{table*}[t]
%\small
%\centering
\caption{Lensing parameters of solutions for KMT-2023-BLG-0830.  \label{table:two}}
\begin{tabular}{lllllcc}
%\begin{tabular}{\columnwidth}{@{\extracolsep{\fill}}lllcc}
\hline\hline
\multicolumn{1}{c}{Parameter}  &
\multicolumn{2}{c}{2L1S}   &
\multicolumn{1}{c}{1L2S}       \\
\multicolumn{1}{c}{}           &
\multicolumn{1}{c}{Close}      &
\multicolumn{1}{c}{Wide }      &
\multicolumn{1}{c}{}           \\
\hline
 $\chi^2$                  &  $1960.2            $    &   $1960.0            $   &  $1975.7            $   \\
 $t_0$ (HJD$^\prime$)      &  $82.2810 \pm 0.0023$    &   $82.2806 \pm 0.0023$   &  $82.2714 \pm 0.0020$   \\
 $u_0$ ($10^{-2}$)         &  $2.387 \pm 0.098   $    &   $0.023 \pm 0.098   $   &  $2.909 \pm 0.183   $   \\
 $\te$ (days)              &  $10.65 \pm 0.40    $    &   $10.89 \pm 0.38    $   &  $10.52 \pm 0.37 1  $   \\
 $s$                       &  $0.931 \pm 0.012   $    &   $1.103 \pm 0.015   $   &   --                    \\
 $q$ ($10^{-3}$)           &  $0.251 \pm 0.096   $    &   $0.237 \pm 0.113   $   &   --                    \\
 $\alpha$ (rad)            &  $4.1900 \pm 0.0065 $    &   $4.1888 \pm 0.0064 $   &   --                    \\
 $\rho$ ($10^{-2}$)        &   --                     &    --                    &  $2.72 \pm 0.42     $   \\
 $t_{0,2}$ (HJD$^\prime$)  &   --                     &    --                    &  $82.4273 \pm 0.0052$   \\
 $u_{0,2}$ ($10^{-3}$)     &   --                     &    --                    &  $-0.17 \pm 0.48    $   \\
 $q_F$                     &   --                     &    --                    &  $0.0029 \pm 0.0016 $   \\
\hline
\end{tabular}
%\tablefoot{  $\chi^2({\rm 1L1S}) = 2022.1$.  }
\end{table*}
% --------------------------------------------------------

The four insets in the bottom panel of Figure~\ref{fig:one} depict the lens system 
configurations, illustrating the trajectory of the source with respect to the caustic 
induced by the planet.  In all degenerate solutions, the planet generates 
a very small central caustic, with a size less than 0.4\% of the Einstein radius. As a result, 
the anomaly lasted for a very short duration. The size of the central caustic depends on both 
the planetary separation $s$ and the mass ratio $q$, shrinking as $s$ deviates from unity and 
as $q$ decreases \citep{Chung2005}. Given the estimated mass ratio of $q \sim 4 \times 10^{-3}$, 
which corresponds to that of a giant planet orbiting a normal star, the short duration of the 
anomaly is attributed to the significant deviation of $s$ from unity. In all solutions, the 
size and shape of the induced caustic are similar. However, for the Local A solutions, the 
source crosses the planet-star axis at an angle of about 41°, while for the Local B solutions, 
the angle is smaller, around $9^\circ$.  Because the source is relatively large compared to the 
caustic, significant finite-source effects occur. Consequently, even though the source passes 
through the caustic, the resulting distortion is relatively small due to these substantial 
finite-source effects \citep{Bennett1996}.

We also considered a model under a 1L2S configuration. The best-fit parameters for the 1L2S 
model are listed in Table~\ref{table:one}, and the corresponding model curve, along with its 
residuals, is shown in Figure~\ref{fig:one}.  From the comparison of fits, it was found that 
the 1L2S model provides a poorer fit compared to the 2L1S model, with a $\chi^2$ difference 
of 12.6, indicating that the 1L2S solution is less favored relative to the 2L1S model.

% Figure 2 ------------------------------------------------------
\begin{figure}[t]
\includegraphics[width=\columnwidth]{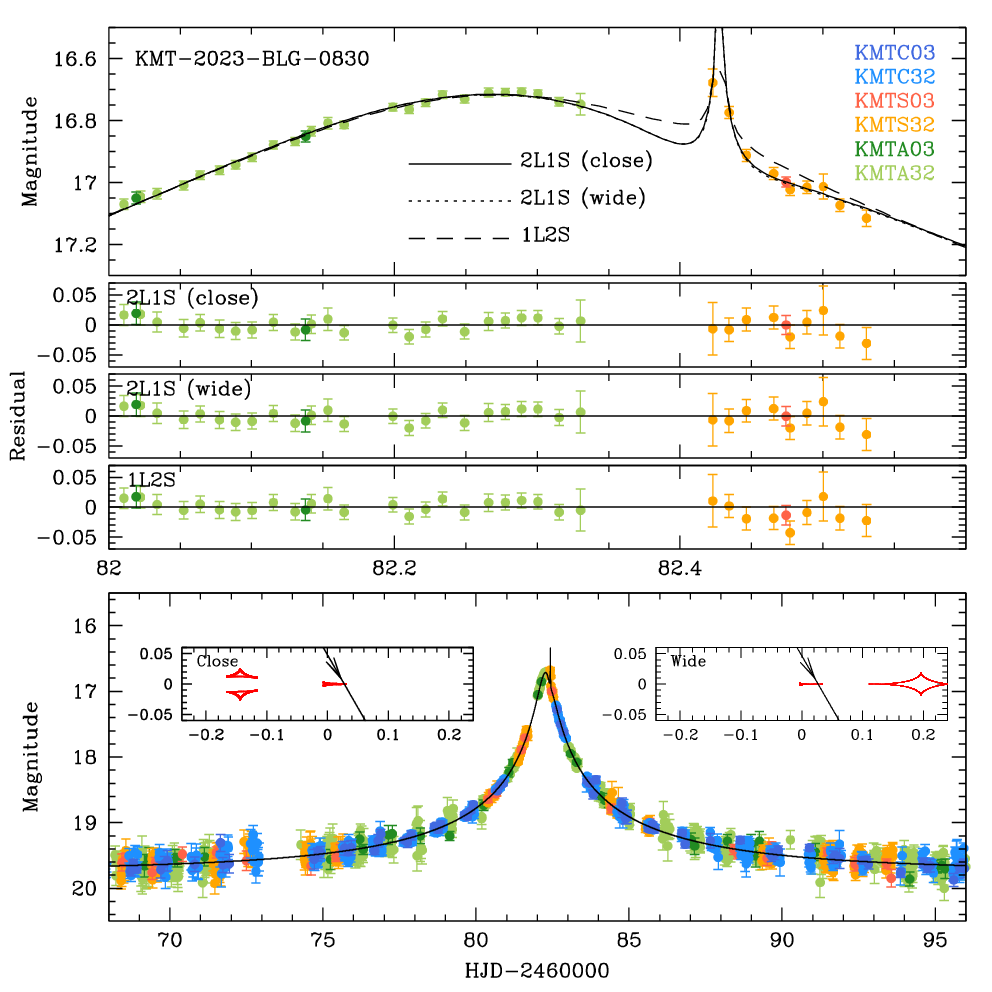}
\caption{
Lensing light curve of KMT-2023-BLG-0830.  The notations are consistent with those in 
Fig.~\ref{fig:one}.
}
\label{fig:two}
\end{figure}
% --------------------------------------------------------------

\subsection{KMT-2023-BLG-0830} \label{sec:four-two}

The source of the event KMT-2023-BLG-0830 is located at $({\rm RA}, {\rm DEC})_{\rm J2000} = 
(18$:06:37.91, -28:19:56.32), corresponding to Galactic coordinates $(l, b) = (2^\circ\hskip-2pt 
.8407$, -3$^\circ \hskip-2pt .6823)$. The source has a baseline $I$-band magnitude of $I_{\rm base} 
= 20.40$, with an extinction of $A_I = 1.01$ in this direction. The event was exclusively observed 
by the KMTNet survey, which first detected the lensing-induced magnification on 2023 May 15 
(${\rm HJD}^\prime = 79$), approximately three days before the peak. The source lies in the 
overlapping region of the BLG03 and BLG32 fields, which were monitored with cadences of 0.5 
hours and 2.5 hours, respectively.

Figure~\ref{fig:two} presents the lensing light curve of KMT-2023-BLG-0830, which, at first 
glance, appears to correspond to a standard 1L1S event with a high peak magnification of 
approximately $A_{\rm max}\sim 50$. However, a detailed examination of the peak region of 
the light curve revealed the presence of a subtle anomaly occurred at around ${\rm HJD}^\prime 
\sim 82.42$.  This anomaly, which has a very short duration of about 1.2 hours, was covered 
by the data obtained from KMTC observations.

Given the abrupt change in lensing magnification during the anomaly and its occurrence near 
the peak of the light curve, we modeled the event under a 2L1S configuration. This analysis 
identified two degenerate solutions resulting from the close-wide degeneracy. From the modeling, 
we estimated the mass ratio between the lens components to be approximately $q \sim 0.25 \times 
10^{-3}$, indicating that the companion to the lens is a low-mass planetary object.  The full 
sets of lensing parameters for both solutions, along with their corresponding $\chi^2$ values, 
are presented in Table~\ref{table:two}.  The degeneracy between the two solutions is very severe, 
with only a marginal difference of $\Delta\chi^2 = 0.2$. As we discuss below, the anomaly was 
caused by the source crossing a central caustic induced by the planet. However, due to the 
limited coverage of the anomaly, the normalized source radius could not be securely determined.

Figure~\ref{fig:two} presents the model light curves and residuals for the two degenerate 
2L1S solutions, while the insets in the bottom panel illustrate the corresponding lens system 
configurations. The configurations reveal that the planet generated a small central caustic 
that is elongated along the planet-host axis, and the anomaly occurred as the source passed 
diagonally through this axis. The short duration of the anomaly resulted from a combination 
of factors, including the small caustic size, which is primarily due to the low planetary mass 
ratio, and the elongated shape of the caustic, which reduced the cross-sectional area traversed 
by the source. In the close solution, the peripheral caustic, which is substantially larger 
than the central caustic, forms on the opposite side of the planet, whereas in the wide solution, 
it appears on the same side as the planet. However, these peripheral caustics are located at a 
considerable distance from the source trajectory and have no impact on the planetary perturbation.

We found that the 1L2S interpretation is disfavored relative to the planetary interpretation. 
This is illustrated in Figure~\ref{fig:two}, which displays the model derived under the 1L2S 
configuration along with its residuals. The lensing parameters for the 1L2S solution are 
detailed in Table~\ref{table:two}. The analysis reveals that the 1L2S model provides a worse 
fit than the planetary solution, with a $\Delta\chi^2=15.7$. Upon examining the residuals, it 
is evident that the 1L2S model fails to adequately account for the negative deviation observed 
after the main anomaly feature.

% Figure 3 ------------------------------------------------------
\begin{figure}[t]
\includegraphics[width=\columnwidth]{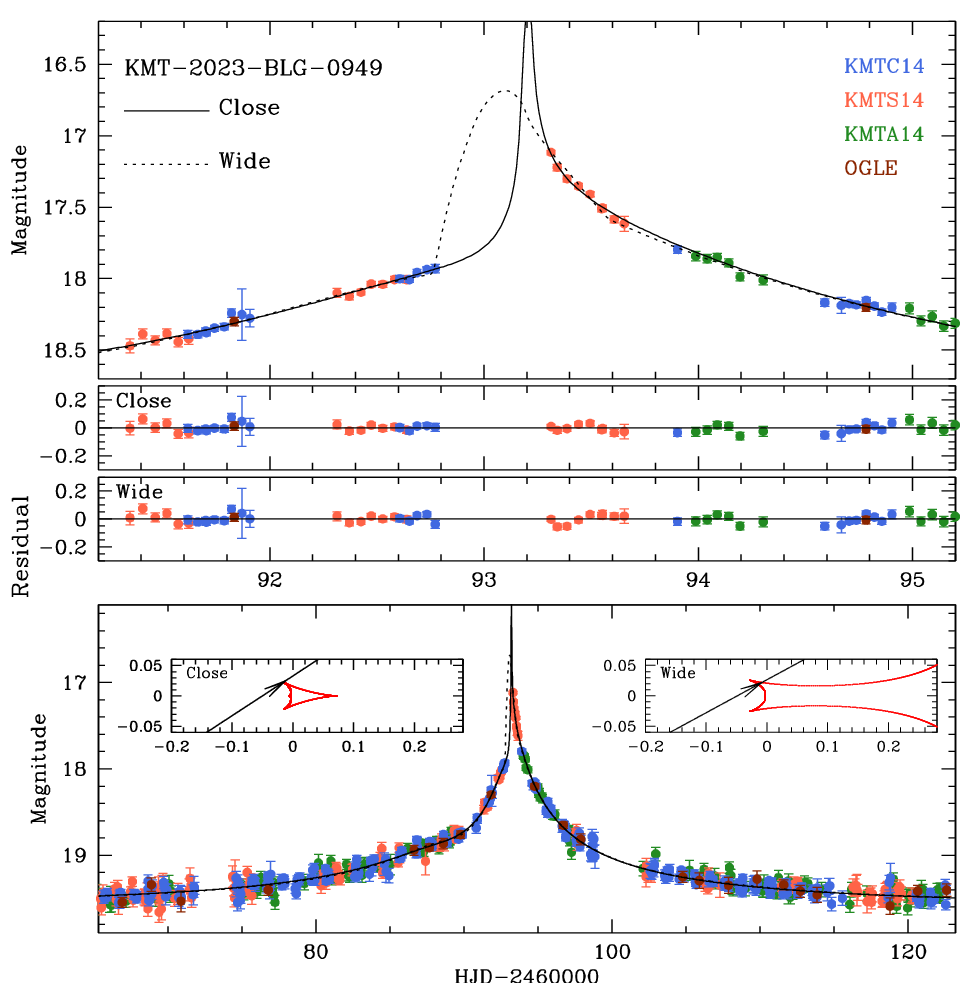}
\caption{
Light curve of KMT-2023-BLG-0949.  We note that in addition to the strong anomaly near the peak, 
there are also additional weak extended negative deviations in the rising part of the light curve.
}
\label{fig:three}
\end{figure}
% --------------------------------------------------------------

\subsection{KMT-2023-BLG-0949} \label{sec:four-three}

The lensing event KMT-2023-BLG-0949 was observed not only by the KMTNet survey but also
through additional observations conducted by the OGLE group. The KMTNet survey first detected
the event on 2023 April 24 (${\rm HJD}^\prime = 86$) during its rising phase, while the OGLE
survey reported the discovery of the event a week later. The OGLE ID reference of the event is
OGLE-2023-BLG-0581.  Hereafter, we refer to the event using the KMTNet ID, following the 
microlensing community's convention of adopting the designation from the survey group that 
first discovered the event.  The equatorial and Galactic coordinates of the event are $({\rm RA}, 
{\rm DEC})_{\rm J2000} = $ (17:33:40.66, \mbox{-28:05:29.26}) and $(l, b) =$ (-$0^\circ\hskip-2pt .6798, 
2^\circ\hskip-2pt .67087)$, respectively.  The baseline magnitude of the source is $I_{\rm base} 
= 19.6$, and the extinction toward the field is $A_I = 2.72$. The event peaked at ${\rm HJD}^\prime 
\sim 93.2$, reaching a relatively high magnification of $A_{\rm peak} \sim 40$.  The source is 
located in the KMTNet BLG14 field, where observations were carried out with an hourly cadence.

Figure~\ref{fig:three} presents the lensing light curve of KMT-2023-BLG-0949, which exhibits 
an anomaly near the peak with a duration of less than a day. The main portion of the anomaly
was covered by the KMTC data, while its declining phase was captured by the KMTA data. A 
detailed examination of the residuals from a 1L1S model revealed the presence of additional 
weak negative deviations in an extended region centered at around ${\rm HJD}^\prime = 90$.

% Table 3 ------------------------------------------------
\begin{table}[t]
%\footnotesize
%\small
%\centering
\caption{Lensing parameters KMT-2023-BLG-0949.\label{table:three}}
%\begin{tabular}{lllllll}
\begin{tabular*}{\columnwidth}{@{\extracolsep{\fill}}lllll}
\hline\hline
\multicolumn{1}{c}{Parameter}        &
\multicolumn{1}{c}{Close}            &
\multicolumn{1}{c}{Wide}            \\
\hline
 $\chi^2$               &  $1710.3           $   &  $1726.0           $  \\  
 $t_0$ (HJD$^\prime$)   &  $93.227 \pm 0.012 $   &  $93.143 \pm 0.016 $  \\
 $u_0$                  &  $0.0274 \pm 0.0024$   &  $0.0238 \pm 0.0019$  \\
 $\te$ (days)           &  $39.19 \pm 2.59   $   &  $42.28 \pm 2.74   $  \\
 $s$                    &  $0.729 \pm 0.010  $   &  $1.184 \pm 0.010  $  \\
 $q$  (10$^{-3}$)       &  $11.77 \pm 1.02   $   &  $7.69 \pm 0.65    $  \\
 $\alpha$ (rad)         &  $2.5624 \pm 0.0102$   &  $2.6413 \pm 0.0095$  \\
 $\rho$ (10$^{-3}$)     &  $< 3              $   &  $< 3              $  \\
\hline             
\end{tabular*}
%\tablefoot{  $\chi^2({\rm 1L1S}) = 2046.7$.}
\end{table}
% -------------------------------------------------------

% Table 4 ------------------------------------------------
\begin{table*}[t]
%\small
%\centering
\caption{Lensing parameters of solutions for KMT-2024-BLG-1281.  \label{table:four}}
\begin{tabular}{lllllcc}
%\begin{tabular}{\columnwidth}{@{\extracolsep{\fill}}lllcc}
\hline\hline
\multicolumn{1}{c}{Parameter}      &
\multicolumn{1}{c}{Inner}          &
\multicolumn{1}{c}{Outer}          &
\multicolumn{1}{c}{Intermediate}   \\
\hline
 $\chi^2$                  &  $677.8            $    &   $678.3            $   &   $666.1            $  \\
 $t_0$ (HJD$^\prime$)      &  $480.891 \pm 0.024$    &   $480.887 \pm 0.023$   &   $480.904 \pm 0.022$  \\
 $u_0$ ($10^{-2}$)         &  $0.0385 \pm 0.0032$    &   $0.0369 \pm 0.0031$   &   $0.0418 \pm 0.0036$  \\
 $\te$ (days)              &  $35.70 \pm 2.49   $    &   $37.27 \pm 2.43   $   &   $33.95 \pm 2.44   $  \\
 $s$                       &  $0.9477 \pm 0.0091$    &   $1.0175 \pm 0.0104$   &   $0.9817 \pm 0.0022$  \\
 $q$ ($10^{-4}$)           &  $2.87 \pm 0.48    $    &   $2.84 \pm 0.51    $   &   $1.75 \pm 0.25    $  \\
 $\alpha$ (rad)            &  $1.396 \pm 0.026  $    &   $1.395 \pm 0.023  $   &   $1.414 \pm 0.022  $  \\
 $\rho$ ($10^{-3}$)        &   --                    &    --                   &    --                  \\
\hline
\end{tabular}
%\tablefoot{  $\chi^2({\rm 1L1S}) = 2022.1$.  }
\end{table*}
% --------------------------------------------------------

From the 2L1S modeling of the light curve, we identified two solutions resulting from the
close-wide degeneracy. The complete lensing parameters for these solutions are presented in
Table~\ref{table:three}. The estimated mass ratio between the lens components is $q = (11.77 
\pm 1.02) \times 10^{-3}$ for the close solution and $q = (7.69 \pm 0.65) \times 10^{-3}$ 
for the wide solution, indicating that the companion to the lens is a giant planet. The model 
curves for the two solutions in the region around the anomaly are shown in the top panel of 
Figure~\ref{fig:three}.  It is found that the close solution is preferred over the wide solution 
by $\Delta\chi^2 = 15.7$.  A notable point is that the models in the region not covered by the 
data are substantially different, suggesting that the degeneracy between the two solutions 
could have been resolved if the sky at the KMTA sites on the night of ${\rm HJD}^\prime = 93$ 
(2023 May 28) had not been clouded out.  Due to the partial coverage of the anomaly, the 
normalized source radius could not be securely constrained, and only its upper limit 
($\rho_{\rm max} \sim 3 \times 10^{-3}$) is determined. Modeling the anomaly under a 1L2S 
configuration results in a solution that is significantly worse than the planetary solution, 
with a $\Delta\chi^2$ of 244.2.

The lens-system configurations for the close and wide solutions are displayed in the two 
insets within the bottom panel of Figure~\ref{fig:three}.  Unlike previous events, 
KMT-2023-BLG-0949 reveals that the central caustics of the close and wide solutions exhibit 
relatively less similarity. In the close solution, the central caustic and the peripheral 
caustic are clearly separated, whereas in the wide solution, the two caustics merge to form 
a single large resonant caustic. In both cases, the anomaly occurred when the source crossed 
the upper left tip of the central caustic. Despite the relatively large size of the caustic, 
the anomaly had a short duration because the source traversed only a small portion of it. 
Additionally, the weak negative deviation observed before the peak resulted from the source 
passing through the extended negative deviation region on the back side of the central caustic. 
Similar planetary signals with extended weak negative deviations have been observed in four 
previously reported planetary lensing events: KMT-2020-BLG-0757, KMT-2022-BLG-0732, 
KMT-2022-BLG-1787, and KMT-2022-BLG-1852 \citep{Han2024a}.

\subsection{KMT-2024-BLG-1281} \label{sec:four-four}

The lensing event KMT-2024-BLG-1281 occurred on a faint source with an $I$-band baseline 
magnitude of $I_{\rm base} = 19.70$. It was first identified by the KMTNet survey on June 
4, 2024 (${\rm HJD}^\prime = 465$), during the rising part of the light curve, and was later 
confirmed by the MOA survey, which designated the event as MOA-2024-BLG-077. The equatorial 
and Galactic coordinates of the event are (RA, DEC)$_{\rm J2000} =$ (11:00:40.65, \mbox{-32:42:51.30}) 
and $(l, b) = $(-1$^\circ\hskip-2pt $.6220, -$4^\circ\hskip-2pt .7000)$, respectively. The event 
lies within the KMTNet BLG34 field, which is characterized by relatively low extinction ($A_I = 
1.04$) compared to other KMTNet fields. This field was monitored at a cadence of 2.5 hours. 
The event reached a relatively high magnification of $A_{\rm max} \sim 25$ at ${\rm HJD}^\prime 
\sim 480.9$.

% Figure 4 ------------------------------------------------------
\begin{figure}[t]
\includegraphics[width=\columnwidth]{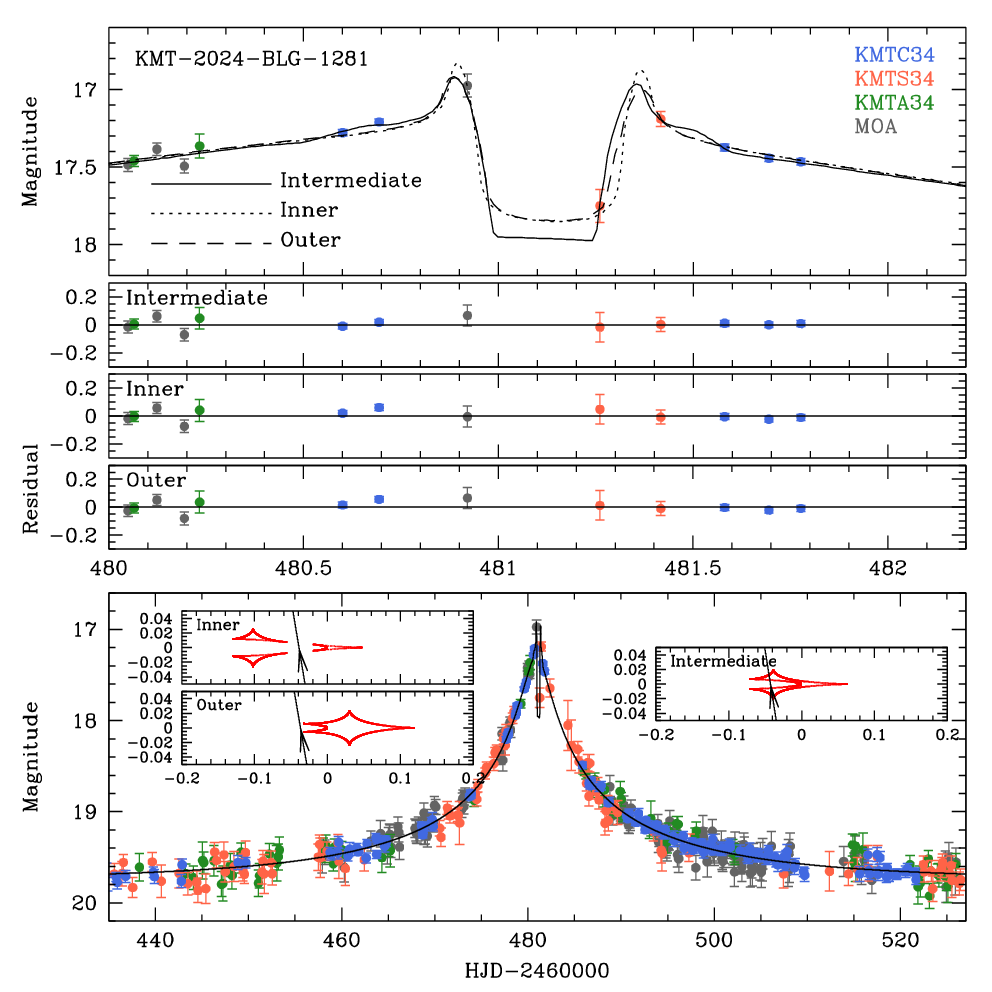}
\caption{
Light curve of the lensing event KMT-2024-BLG-1281.
}
\label{fig:four}
\end{figure}
% --------------------------------------------------------------

The peak region of a high-magnification event is sensitive to perturbations caused by a 
planetary companion. Upon examining the light curve with this in mind, we identified a 
short-lived anomaly lasting less than a day, centered around ${\rm HJD}^\prime \sim 481.2$.  
Figure~\ref{fig:four} presents the light curve of KMT-2024-BLG-1281, with the lower panel 
displaying the full view and the upper panel providing a zoomed-in view of the anomaly 
region. One KMTS data point at ${\rm HJD}^\prime = 481.269$ shows a negative deviation 
from the 1L1S model, while two other points -- a MOA point at ${\rm HJD}^\prime = 480.927$ 
and a KMTS point at ${\rm HJD}^\prime = 481.423$ -- exhibit positive deviations relative 
to the 1L1S model.  Besides these points, the KMTC point at ${\rm HJD}^\prime = 60480.698$ 
appears to slightly deviate from the 1L1S model.

Light curve modeling under the 2L1S configuration revealed three degenerate solution sets. 
In all cases, the mass ratio between the lens components is approximately $q \sim 2 \times 
10^{-4}$, indicating that the companion is planetary in nature. For each solution, the 
normalized separation between the lens components is close to unity, implying that the 
planet resides near the Einstein ring. The complete set of lensing parameters for each 
solution is listed in Table~\ref{table:four}.  The solutions are referred to as “inner,” 
“outer,” and “intermediate,” as described in the discussion below.  The top panel of 
Figure~\ref{fig:four} presents the model light curves for the three solutions, along with 
their corresponding residuals. The intermediate solution yields a better fit than the inner 
and outer solutions, with improvements of $\Delta\chi^2 = 11.7$ and 12.2, respectively.  
The normalized source radius could not be constrained because the anomaly was sparsely 
covered.

The insets within the bottom panel of Figure~\ref{fig:four} displays the lens-system 
configurations corresponding to the three solutions. According 
to the inner solution, the source passed through the inner side of the peripheral caustic 
relative to the planet host, while for the outer solution, the source trajectory lay on 
the outer side of the caustic. In the intermediate solution, the source crossed directly 
over the caustic. Based on these configurations, we label the solutions as ``inner,'' ``outer,'' 
and ``intermediate,'' respectively.  The degeneracy between the inner and outer solutions was 
first identified by \citet{Gaudi1997}, who described it as arising from source trajectories 
that pass on opposite sides -- inner and outer -- of a planetary caustic. \citet{Yee2021} 
and \citet{Zhang2022} later demonstrated that this type of degeneracy can also occur in 
planetary perturbations induced by both central and planetary caustics. Subsequently, 
\citet{Hwang2022} and \citet{Gould2022} proposed analytic relations linking the planetary 
parameters of the degenerate solutions.  For all solutions, the source trajectory passes 
nearly vertically through the perturbation region, which is elongated along the planet-host 
axis. Combined with the small caustic size generated by the low-mass planet, this results 
in a short-duration perturbation.

\subsection{KMT-2024-BLG-2059} \label{sec:four-five}

The microlensing event KMT-2024-BLG-2059 occurred on a source with a baseline magnitude 
of $I_{\rm base} = 18.12$, located at equatorial coordinates (RA, DEC)$_{\rm J2000}$ = 
(18:03:43.14, \mbox{-28:34:05.48}), corresponding to Galactic coordinates $(l, b) = (2^\circ
\hskip-2pt.3220$, -$3^\circ\hskip-2pt .2378)$.  The event was first detected by the KMTNet 
survey on 2024 August 5 (${\rm HJD}^\prime = 527$), two days prior to its peak. It was 
later independently identified by the OGLE survey on August 16 (${\rm HJD}^\prime = 538$) 
and designated as OGLE-2024-BLG-1095. The source lies in the overlapping region of the two 
KMTNet prime fields, BLG03 and BLG43, which were monitored with a combined cadence of 
0.25 hours. This region is located near Baade’s Window, where the extinction is relatively 
low ($A_I = 0.96$).

\begin{figure}[t]
\includegraphics[width=\columnwidth]{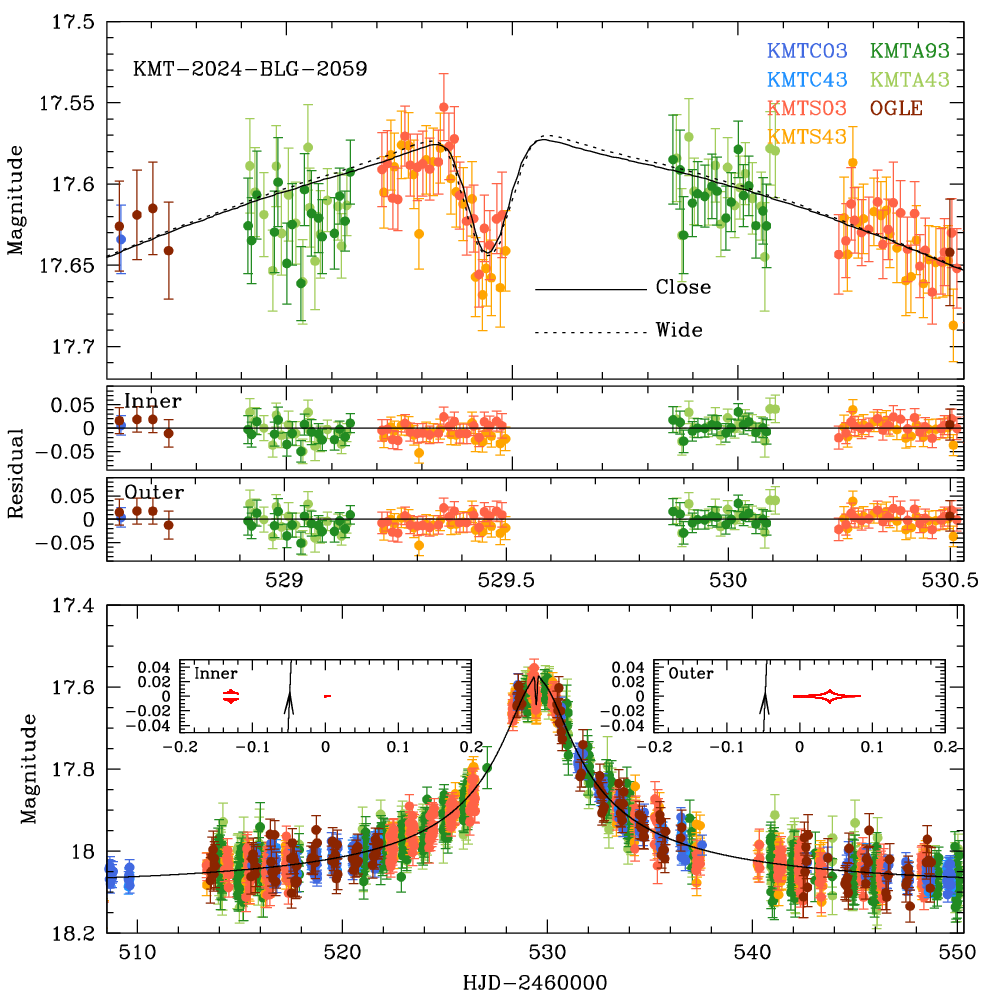}
\caption{
Lensing light curve of the event KMT-2024-BLG-2059.
}
\label{fig:five}
\end{figure}

Figure~\ref{fig:five} shows the lensing light curve of the event.  Modeling under the 
1L1S configuration revealed that the observed flux is substantially influenced by blended 
light.  As a result, although the event reached a relatively high magnification of 
$A_{\rm max} \sim 20$, the source brightened by only $\sim 0.52$ magnitudes above the 
baseline at the peak. Due to the sensitivity of high-magnification events to planetary 
perturbations, we examined the peak region of the light curve.  This examination revealed 
a brief anomaly lasting about 9 hours, as displayed in the top panel of Figure~\ref{fig:five}.  
The anomaly, captured by two KMTS data sets, displays a negative deviation from the 1L1S model. 
As demonstrated by the recently reported planetary events MOA-2022-BLG-033Lb, KMT-2023-BLG-0119Lb, 
and KMT-2023-BLG-1896Lb \citep{Han2025}, such short-duration dips near the peak of a 
high-magnification event are strong indicators of a planetary origin.

% Figure 5 ------------------------------------------------------
% --------------------------------------------------------------

% Table 5 ------------------------------------------------
\begin{table}[t]
%\footnotesize
%\small
%\centering
\caption{Lensing parameters KMT-2024-BLG-2059.\label{table:five}}
%\begin{tabular}{lllllll}
\begin{tabular*}{\columnwidth}{@{\extracolsep{\fill}}lllll}
\hline\hline
\multicolumn{1}{c}{Parameter}        &
\multicolumn{1}{c}{Inner}            &
\multicolumn{1}{c}{Outer}            \\
\hline
 $\chi^2$               &  $8599.50          $   &  $8599.27          $  \\  
 $t_0$ (HJD$^\prime$)   &  $529.512 \pm 0.016$   &  $529.519 \pm 0.016$  \\
 $u_0$                  &  $0.0480 \pm 0.0030$   &  $0.0472 \pm 0.0030$  \\
 $\te$ (days)           &  $31.68 \pm 1.71   $   &  $31.91 \pm 1.72   $  \\
 $s$                    &  $0.937 \pm 0.014  $   &  $1.021 \pm 0.015  $  \\
 $q$  (10$^{-5}$)       &  $3.90 \pm 2.39    $   &  $4.04 \pm 2.51    $  \\
 $\alpha$ (rad)         &  $1.607 \pm 0.012  $   &  $1.616 \pm 0.012  $  \\
 $\rho$ (10$^{-3}$)     &   --                   &   --                  \\
\hline             
\end{tabular*}
%\tablefoot{  $\chi^2({\rm 1L1S}) = 2046.7$.}
\end{table}
% -------------------------------------------------------

% Table 6 ------------------------------------------------
\begin{table*}[t]
%\small
%\centering
\caption{Lensing parameters of KMT-2024-BLG-2242.  \label{table:six}}
\begin{tabular}{lllllcc}
%\begin{tabular}{\columnwidth}{@{\extracolsep{\fill}}lllcc}
\hline\hline
\multicolumn{1}{c}{Parameter}  &
\multicolumn{2}{c}{2L1S}       &
\multicolumn{1}{c}{1L2S}       \\
\multicolumn{1}{c}{}           &
\multicolumn{1}{c}{Inner}      &
\multicolumn{1}{c}{Outer}      &
\multicolumn{1}{c}{}           \\
\hline
 $\chi^2$                  &  $9514.4              $   &   $9510.2             $   &  $9519.7             $   \\
 $t_0$ (HJD$^\prime$)      &  $545.2133 \pm 0.0044 $   &   $545.2108 \pm 0.0046$   &  $545.2040 \pm 0.0046$   \\
 $u_0$                     &  $0.480 \pm 0.011     $   &   $0.480 \pm 0.011    $   &  $0.507 \pm 0.013    $   \\
 $\te$ (days)              &  $4.687 \pm 0.068     $   &   $4.692 \pm 0.067    $   &  $4.596 \pm 0.072    $   \\
 $s$                       &  $1.384 \pm 0.031     $   &   $1.179 \pm 0.026    $   &   --                     \\
 $q$ ($10^{-4}$)           &  $7.52 \pm 1.511      $   &   $6.04 \pm 1.51      $   &   --                     \\
 $\alpha$ (rad)            &  $4.6308 \pm 0.0051   $   &   $4.6223 \pm 0.0051  $   &   --                     \\
 $\rho$ ($10^{-3}$)        &  $  41.64 \pm 11.18   $   &   $43.22 \pm 10.56    $   &   --                     \\
 $t_{0,2}$ (HJD$^\prime$)  &    --                     &    --                     &  $545.4071 \pm 0.0095$   \\
 $u_{0,2}$                 &    --                     &    --                     &  $-0.003 \pm 0.012   $   \\
 $\rho_2$ ($10^{-3}$)      &    --                     &    --                     &  $30.21 \pm 3.42     $   \\
 $q_F$ ($10^{-3}$)         &    --                     &    --                     &  $2.01 \pm 0.31      $   \\
\hline
\end{tabular}
%\tablefoot{  $\chi^2({\rm 1L1S}) = 2022.1$.  }
\end{table*}
% --------------------------------------------------------

Detailed modeling of the light curve confirmed the planetary nature of the anomaly.  
The analysis yielded two local solutions arising from the inner-outer degeneracy. 
The planet parameters are $(s, q) \sim (0.94, 3.9\times 10^{-5})$ for the inner 
solution and $(1.02, 4.0\times 10^{-5})$ for the outer solution. The full sets of 
lensing parameters for both solutions are listed in Table~\ref{table:five}.  We exclude 
the 1L2S interpretation, as perturbations induced by a source companion always produce 
positive deviations. The model curves for the two solutions, along with their residuals, 
are presented in the upper panels of Figure~\ref{fig:five}.  The fits are nearly 
indistinguishable, with a $\chi^2$ difference of only $\Delta\chi^2 = 0.23$.

The two insets in the bottom panel of Figure~\ref{fig:five} illustrate the lens-system 
configurations corresponding to the inner and outer solutions. As in the case of 
KMT-2024-BLG-1281, the source trajectory passes through the inner side of the 
peripheral caustic in the inner solution and through the outer side in the outer 
solution. The dip-like anomaly arose as the source crossed a region of negative 
deviation located on the backside of the central caustic along the planet–host axis.  
The short duration of the perturbation results from the nearly vertical source trajectory 
across the anomaly region, combined with the small caustic size produced by the low-mass 
planet. The normalized source radius could not be constrained, as the source did not 
intersect the caustic.

\subsection{KMT-2024-BLG-2242} \label{sec:four-six}

The lensing event KMT-2024-BLG-2242 occurred on a source with a relatively bright 
baseline magnitude of $I_{\rm base} = 16.68$. The source is located at equatorial 
coordinates (RA, DEC)$_{\rm J2000}$ = (17:53:39.60, -29:33:37.12) and Galactic 
coordinates $(l, b) = (0^\circ\hskip-2pt .3655$, -1$^\circ\hskip-2pt .8252)$. The 
extinction toward this field is $A_I = 1.84$. The event was independently identified 
in its early phase by all three active microlensing surveys. KMTNet and MOA detected it 
on the same day, August 19 (${\rm HJD}^\prime = 541$), while OGLE reported the detection 
several days later. We adopt the KMTNet designation for the event. The corresponding event 
IDs from the MOA and OGLE surveys are MOA-2024-BLG-204 and OGLE-2024-BLG-1122, respectively. 
The event reached a peak magnification of $A_{\rm max} \sim 2.3$ at ${\rm HJD}^\prime = 
545.2$. Although the timescale was short, $\te \sim 4.6$ days as derived from a 1L1S model 
fit, the light curve was densely covered thanks to the event’s location in KMTNet prime 
fields BLG02 and BLG42, which were monitored with a combined cadence of 15 minutes.

% Figure 6 ------------------------------------------------------
\begin{figure}[t]
\includegraphics[width=\columnwidth]{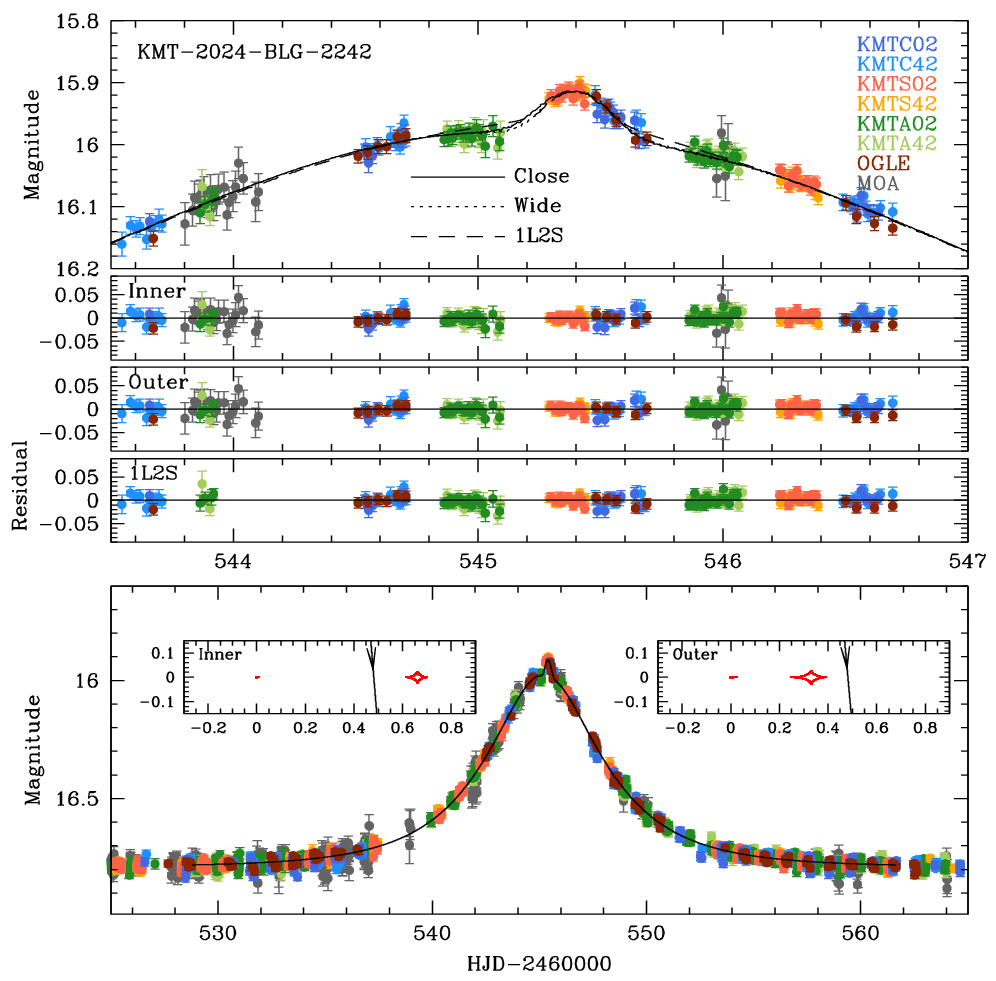}
\caption{
Lensing light curve of KMT-2024-BLG-2242.
}
\label{fig:six}
\end{figure}
% --------------------------------------------------------------

The lensing light curve of KMT-2024-BLG-2242 is shown in Figure~\ref{fig:six}.  Although 
it initially resembles a standard 1L1S event, a closer inspection of the peak region reveals 
a brief anomaly lasting less than a day. This deviation, captured by the combined KMTS, KMTC, 
and OGLE data sets, appears as a bump with a positive deviation from the 1L1S model. If this 
anomaly is attributed to a planetary companion, its short duration suggests the presence of 
a low-mass planet. In such a case, it is unlikely that the perturbation was caused by the 
central caustic, as the central caustic induced by a low-mass planet is too small to generate 
a signal at low magnification. A more plausible interpretation is that the bump-like anomaly 
resulted from the source passing nearly perpendicularly through the perturbation region 
associated with the peripheral caustic.

% Figure 7 ------------------------------------------------------
\begin{figure*}[t]
\centering
\includegraphics[width=18.3cm]{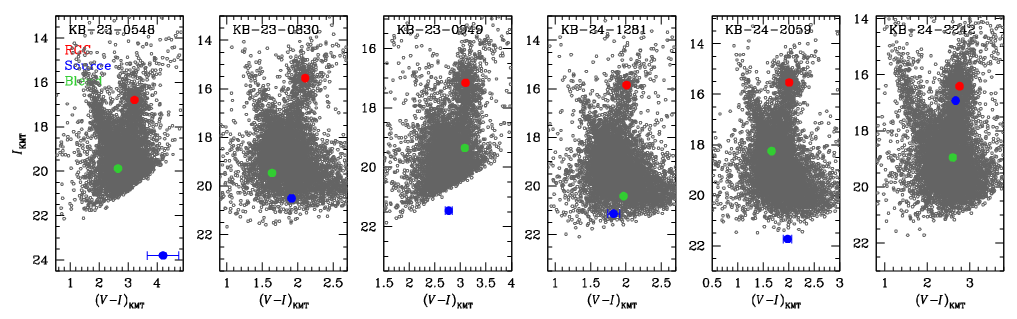}
\caption{
Locations of source stars (blue dots) with respect to the centroid of the red giant clump 
(RGC, red dots) in the instrumental color-magnitude diagrams of the six planetary lensing 
events.  For events with measured blended flux, the positions of the blend (green dots) 
are also indicated.
}
\label{fig:seven}
\end{figure*}
% --------------------------------------------------------------

% Table 7 ------------------------------------------------
\begin{table*}[t]
\small
%\centering
\caption{Source parameters. \label{table:seven}}
\begin{tabular}{lcccccccclll}
%\begin{tabular}{\columnwidth}{@{\extracolsep{\fill}}lllcc}
\hline\hline
\multicolumn{1}{c}{Event}              &
\multicolumn{1}{c}{$(V-I)_{s}$ }  &
\multicolumn{1}{c}{$(V-I)_{\rm RGC}$ }  &
\multicolumn{1}{c}{$(V-I)_{s,0}$ }  &
\multicolumn{1}{c}{$I_{s}$ }  &
\multicolumn{1}{c}{$I_{\rm RGC}$ }  &
\multicolumn{1}{c}{$I_{s,0}$ }  &
\multicolumn{1}{c}{$\theta_*$ ($\mu$as)   } \\
\hline
KMT-2023-BLG-0548  & $4.203 \pm 0.543$  &  3.218  &  $2.045 \pm 0.545$   &  $23.797 \pm 0.023$  &  14.343   &  $21.347 \pm 0.030$  & $0.43  \pm 0.24$   \\
KMT-2023-BLG-0830  & $1.913 \pm 0.026$  &  2.105  &  $0.868 \pm 0.048$   &  $20.511 \pm 0.004$  &  14.354   &  $19.300 \pm 0.020$  & $0.518 \pm 0.044$  \\
KMT-2023-BLG-0949  & $2.772 \pm 0.068$  &  3.101  &  $0.731 \pm 0.079$   &  $21.465 \pm 0.005$  &  14.339   &  $18.644 \pm 0.021$  & $0.600 \pm 0.063$  \\   
KMT-2024-BLG-1281  & $1.825 \pm 0.085$  &  2.009  &  $0.876 \pm 0.094$   &  $21.143 \pm 0.008$  &  14.535   &  $19.828 \pm 0.021$  & $0.410 \pm 0.048$  \\   
KMT-2024-BLG-2059  & $1.980 \pm 0.084$  &  2.011  &  $1.029 \pm 0.093$   &  $21.730 \pm 0.010$  &  14.593   &  $20.784 \pm 0.022$  & $0.314 \pm 0.037$  \\   
KMT-2024-BLG-2242  & $2.669 \pm 0.018$  &  2.762  &  $0.967 \pm 0.044$   &  $16.926 \pm 0.004$  &  14.545   &  $15.057 \pm 0.020$  & $4.213 \pm 0.348$  \\
\hline
\end{tabular}
\tablefoot{$(V-I)_{{\rm RGC},0}=1.06$.}
\end{table*}
% --------------------------------------------------------

Detailed modeling confirms the planetary origin of the anomaly. We identified two degenerate 
solutions resulting from the inner-outer degeneracy. The planetary parameters are approximately 
$(s, q) \sim (1.35, 7.5\times 10^{-4})$ for the inner solution and $(s, q) \sim (1.18, 6.0
\times 10^{-4})$ for the outer solution, indicating that the companion is a low-mass planet. 
The complete parameter sets for both solutions are provided in Table~\ref{table:six}, and the 
corresponding model curves and residuals are presented in the upper panels of Figure~\ref{fig:six}.  
The two solutions are very degenerate, with the outer solution being slightly favored by 
$\Delta\chi^2 = 4.2$.

We also considered the possibility that the anomaly was caused by a binary companion to the 
source. Modeling under a 1L2S configuration produced a fit comparable to those of the planetary 
models. The full set of parameters for the 1L2S solution is presented in Table~\ref{table:six}. 
According to this model, the source companion is extremely faint, with a flux ratio to the 
primary of $q_F \sim 2 \times 10^{-3}$. However, the inferred normalized source radius, $\rho_2 
\sim 30 \times 10^{-3}$, corresponds to that of a giant star. This inconsistency between the 
companion's faintness and its large angular size is physically implausible, leading us to reject 
the 1L2S interpretation for the anomaly.

The lens-system configurations corresponding to the inner and outer planetary solutions are
presented in the two insets within the bottom panel of Figure~\ref{fig:six}.  As anticipated 
from the heuristic analysis, the anomaly was caused by the source passing nearly vertically 
through the perturbation region induced by the peripheral caustic. In the inner solution, the 
source trajectory passes on the inner side of the caustic, while in the outer solution, it 
passes on the outer side.  The short duration of the anomaly is primarily due to the small 
size of the caustic, which results from the low planet-to-host mass ratio. Although the source 
does not cross the caustic in either solution, the normalized source radius was nonetheless 
constrained, albeit with relatively large uncertainty. This was made possible by the substantial 
angular size of the source, as discussed in the following section.

\section{Source stars} \label{sec:five}

Identifying the source of a lensing event is essential not only for fully characterizing the 
event but also for estimating the angular Einstein radius. Determining the source type allows 
one to estimate the angular source radius, $\theta_*$. With this value, the angular Einstein 
radius can be calculated using the relation
\begin{equation}
\theta_{\rm E} = {\theta_* \over \rho},
\label{eq1}
\end{equation}
where $\rho$ is the normalized source radius, derived from modeling events that exhibit 
measurable finite-source effects. Among the analyzed events, $\rho$ was reliably measured 
for KMT-2023-BLG-0548, KMT-2024-BLG-1281, and KMT-2024-BLG-2242, allowing for the estimation 
of the angular Einstein radius $\thetae$.  For KMT-2023-BLG-0949, only an upper limit on 
$\rho$ could be determined, which in turn provides a lower limit on $\thetae$.

To determine the source type, we began by estimating the instrumental color and magnitude. 
This was achieved by regressing the observed flux $F_{\rm obs}(t)$ in the $I$ and $V$ 
passbands against the model magnification $A_{\rm model}(t)$;
\begin{equation}
F_{\rm obs}(t) = A_{\rm model}(t) F_s + F_b,
\label{eq2}
\end{equation}
where $F_s$ and $F_b$ represent the flux values of the source and the blend, respectively.  
The instrumental color and magnitude, $(V-I, I)_s$, were subsequently calibrated using the 
method described by \citet{Yoo2004}, which employs the centroid of the red giant clump (RGC) 
in the color-magnitude diagram (CMD) as a reference. The RGC centroid serves as a reference 
for calibration because its de-reddened color and magnitude, $(V-I, I)_{\rm RGC,0}$, are known 
in previous studies by \citet{Bensby2013} and \citet{Nataf2013}. By measuring the offsets 
$\Delta(V-I, I)$ between the source and the RGC centroid in the instrumental CMD, the 
de-reddened values of the source are determined as:
\begin{equation}
(V - I, I)_{s,0} = (V - I, I)_{{\rm RGC},0} + \Delta(V - I, I).
\label{eq3}
\end{equation}

% Table 8 ------------------------------------------------
\begin{table}[t]
%\footnotesize
%\small
%\centering
\caption{Angular Einstein radii.\label{table:eight}}
%\begin{tabular}{lllllll}
\begin{tabular*}{\columnwidth}{@{\extracolsep{\fill}}lllll}
\hline\hline
\multicolumn{1}{c}{Event}                      &
\multicolumn{1}{c}{$\thetae$ (mas)}            &
\multicolumn{1}{c}{$\mu$ (mas/yr)}            \\
\hline
 KMT-2023-BLG-0548   &  $0.38 \pm 0.23  $   &   $3.91 ±\pm2.38  $   \\  
 KMT-2023-BLG-0949   &  $> 0.2          $   &   $> 1.7          $   \\
 KMT-2024-BLG-1281   &  $0.233 \pm 0.069$   &   $2.260 \pm 0.671$   \\
 KMT-2024-BLG-2242   &  $0.090 \pm 0.021$   &   $6.963 \pm 1.667$   \\
\hline             
\end{tabular*}
%\tablefoot{  $\chi^2({\rm 1L1S}) = 2046.7$.}
\end{table}
% -------------------------------------------------------

Figure~\ref{fig:seven} shows the positions of the source stars relative to the RGC centroids 
on the instrumental CMDs, constructed from stars located near the source in each event.  
For events with measured blended ﬂux, we also indicate the positions of the blend.  The CMDs 
were generated using photometry performed with the pyDIA photometry code \citep{Albrow2017}, 
which was also used to measure the source color.  The estimated values of $(V-I, I)_s$, 
$(V-I, I)_{\rm RGC}$, $(V-I, I)_{\rm RGC,0}$, and $(V-I, I)_{s,0}$ are presented in 
Table~\ref{table:seven}.  With the exception of KMT-2024-BLG-2242, the measured colors and 
magnitudes indicate that the source stars are main-sequence stars with spectral types ranging 
from mid-G to early-M.  In the case of KMT-2024-BLG-2242, the source is identified as a K-type 
giant with a much larger angular radius than those of the other events. This accounts for the 
ability to measure the normalized source radius in this case, even though the source passed 
the caustic at a relatively large separation.

The angular source radius was estimated using the measured color and magnitude of the source 
star, based on the empirical relation of \citet{Kervella2004}, which links $(V-K, I)$ to 
$\theta_*$. Because this relation requires $(V-K, K)$ as input, the observed $(V-I, I)$ 
values were converted to $(V-K, K)$ using the color-color transformation provided by 
\citet{Bessell1988}. With the derived angular source radius, the angular Einstein radius, 
$\thetae$, was computed using Equation~(\ref{eq1}). The relative lens-source proper motion, 
$\mu$, was then obtained from the event timescale, $\te$, determined through modeling, via 
the relation $\mu = \thetae / \te$.

The resulting values of $\thetae$ and $\mu$ are summarized in Table~\ref{table:eight} for 
the three events exhibiting finite-source effects: KMT-2023-BLG-0548, KMT-2024-BLG-1281, 
and KMT-2024-BLG-2242. For KMT-2023-BLG-0949, where only an upper limit on $\rho$ could be 
constrained, lower limits on $\thetae$ and $\mu$ are provided. Notably, the angular Einstein 
radius measured for KMT-2024-BLG-2242, $\thetae = 0.090 \pm 0.021$~mas, is significantly 
smaller than those of the other events. Combined with its short timescale of $\te = 4.687 
\pm 0.068$ days, this indicates that the lens likely has a very low mass.

% Table 9 ------------------------------------------------
\begin{table*}[t]
%\small
%\centering
\caption{Physical lens parameters. \label{table:nine}}
\begin{tabular}{llllllllllll}
%\begin{tabular}{\columnwidth}{@{\extracolsep{\fill}}lllcc}
\hline\hline
\multicolumn{1}{c}{Event}                       &
\multicolumn{4}{c}{Parameters}                  &
\multicolumn{1}{c}{$p_{\rm disk}$ }             &
\multicolumn{1}{c}{$p_{\rm bulge}$ }            \\
\multicolumn{1}{c}{Model}                       &
\multicolumn{1}{c}{$M_{\rm h}$ ($M_\odot$) }    &
\multicolumn{1}{c}{$M_{\rm p}$ ($M_{\rm J}$)}   &
\multicolumn{1}{c}{$\dl$ (kpc) }                &
\multicolumn{1}{c}{$a_\perp$ (AU)}              &
\multicolumn{1}{c}{(\%) }                       &
\multicolumn{1}{c}{(\%) }                       \\
\hline
KMT-2023-BLG-0548        &        \\
Low mass                 &  $0.123^{+0.073}_{-0.056}$ &                             &    $4.32^{+1.85}_{-1.38}$     &                         & 77  &  23   \\  [0.3ex]
\ \ \ Local A, close     &   --                       &  $0.69^{+0.41}_{-0.31}$     &     --                        &  $0.49^{+0.09}_{-0.08}$ &     &       \\  [0.3ex]
\ \ \ Local A, wide      &   --                       &  $0.58^{+0.35}_{-0.27}$     &     --                        &  $2.82^{+0.50}_{-0.46}$ &     &       \\  [0.3ex]
\ \ \ Local B, close     &   --                       &  $0.51^{+0.30}_{-0.23}$     &     --                        &  $0.45^{+0.08}_{-0.07}$ &     &       \\  [0.3ex]
\ \ \ Local B, wide      &   --                       &  $0.58^{+0.35}_{-0.27}$     &     --                        &  $3.22^{+0.58}_{-0.52}$ &     &       \\  [0.3ex]
High mass                &  $0.72^{+0.26}_{-0.19}$    &                             &    $6.83^{+1.22}_{-1.10}$     &                         & 27  &  73   \\  [0.3ex]
\ \ \ Local A, close     &   --                       &  $4.04^{+1.43}_{-1.05}$     &     --                        &  $1.18^{+0.21}_{-0.19}$ &     &       \\  [0.3ex]
\ \ \ Local A, wide      &   --                       &  $3.42^{+1.21}_{-0.89}$     &     --                        &  $6.85^{+1.22}_{-1.10}$ &     &       \\  [0.3ex]
\ \ \ Local B, close     &   --                       &  $3.01^{+1.07}_{-0.79}$     &     --                        &  $1.10^{+0.20}_{-0.18}$ &     &       \\  [0.3ex]
\ \ \ Local B, wide      &   --                       &  $3.45^{+1.22}_{-0.90}$     &     --                        &  $7.81^{+1.40}_{-1.26}$ &     &       \\  [0.3ex]
\hline                                                                                                                                                            
KMT-2023-BLG-0830        &  $0.22^{+0.26}_{-0.13}$    &                             &    $7.25^{+1.13}_{-1.21}$     &                         & 14  &  86   \\  [0.3ex]
\ \ \ Close              &   --                       &  $0.059^{+0.068}_{-0.035}$  &     --                        & $1.43^{+0.22}_{-0.24}$  &     &       \\  [0.3ex]
\ \ \ Wide               &   --                       &  $0.056^{+0.064}_{-0.033}$  &     --                        & $1.70^{+0.27}_{-0.29}$  &     &       \\  [0.3ex]
\hline                                                                                                              
KMT-2023-BLG-0949        &  $0.48^{+0.35}_{-0.25} $   &                             &    $6.06^{+1.64}_{-2.41}$     &                         & 54  &  46   \\  [0.3ex]
\ \ \ Close              &   --                       &  $5.98^{+4.36}_{-3.11}$     &     --                        & $2.03^{+0.66}_{-0.81}$  &     &       \\  [0.3ex]
\ \ \ Wide               &   --                       &  $3.91^{+2.85}_{-2.03}$     &     --                        & $3.30^{+0.89}_{-1.32}$  &     &       \\  [0.3ex]
\hline                                                                                                              
KMT-2024-BLG-1281        &  $0.32^{+0.35}_{-0.19}$    &  $0.060^{+0.066}_{-0.036}$  &    $7.18^{+1.09}_{-1.40}$     & $1.96^{+0.30}_{-0.38}$  & 24  &  76   \\  [0.3ex]
\hline                                                                                                              
KMT-2024-BLG-2059        &  $0.51^{+0.28}_{-0.30}$    &                             &    $6.07^{+1.32}_{-2.11}$     &                         & 45  &  55   \\  [0.3ex]
\ \ \ Inner              &   --                       &  $0.022^{+0.012}_{-0.013}$  &     --                        & $2.78^{+0.60}_{-0.97}$  &     &       \\  [0.3ex]
\ \ \ Outer              &   --                       &   --                        &     --                        & $2.55^{+0.56}_{-0.89}$  &     &       \\  [0.3ex]
\hline                                                                                                              
KMT-2024-BLG-2242        &  $0.069^{+0.120}_{-0.038}$ &                             &    $7.62^{+1.01}_{-1.03}$     &                         & 22  &  78   \\  [0.3ex]
\ \ \ Inner              &   --                       &  $0.055^{+0.095}_{-0.030}$  &     --                        & $1.05^{+0.14}_{-0.14}$  &     &       \\  [0.3ex]
\ \ \ Outer              &   --                       &  $0.044^{+0.076}_{-0.024}$  &     --                        & $0.89^{+0.12}_{-0.12}$  &     &       \\  
\hline
\end{tabular}
\tablefoot{
$M_{\rm p}$ and $M_{\rm h}$ denote the masses of the planet and its host star, respectively. 
$\dl$ indicates the distance, while $a_\perp$ is the projected separation between the planet 
and the host. The quantities $p_{\rm disk}$ and $p_{\rm bulge}$ represent the probabilities 
that the planetary system is located in the disk and the bulge, respectively.
}
\end{table*}
% --------------------------------------------------------

% Figure 8 ------------------------------------------------------
\begin{figure}[t]
\includegraphics[width=\columnwidth]{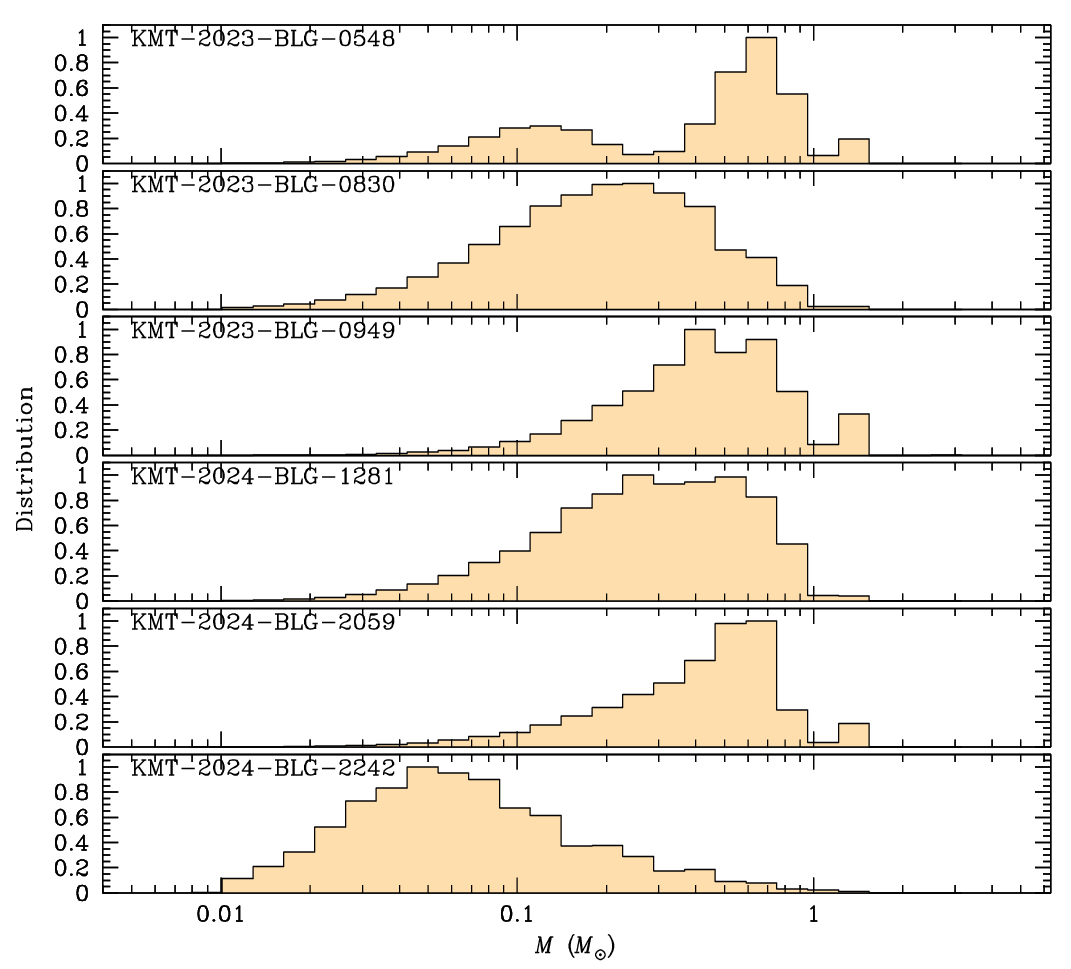}
\caption{
Posteriors for the mass of the lens systems. 
}
\label{fig:eight}
\end{figure}
% --------------------------------------------------------------

\section{Physical lens parameters} \label{sec:six}

The physical parameters of the lens, including its mass ($M$) and distance ($\dl$), 
were determined by performing a Bayesian analysis.  This analysis incorporated priors 
based on the Galactic model and the mass function of Galactic objects, along with 
constraints  provided by the measured lensing observables.  The lensing observables 
that provide constraints on $M$ and $\dl$ are the event timescale $\te$, the angular 
Einstein radius $\thetae$, and the microlens parallax $\pie$ because they are related 
to the physical parameters through the following relations:
\begin{equation}
\te = {\thetae \over \mu}, \quad
\thetae = (\kappa M \pi_{\rm rel})^{1/2}, \quad
\pie = {\pi_{\rm rel} \over \thetae}.
\label{eq4}
\end{equation}
Here $\kappa = 4G/(c^2{\rm AU}) \simeq 8.144~{\rm mas}/M_\odot$,  $\pi_{\rm rel}={\rm AU}
(1/ \ds- 1/\ds)$ represents the relative lens-source parallax, and $\ds$ is the distance 
to the source.  With the complete measurements of these observables, the lens mass and 
distance can be uniquely determined using the relations
\begin{equation}
M = {\thetae\over \kappa \pie};\qquad
\dl = { {\rm AU} \over \pie\thetae + \pi_{\rm S}},
\label{eq5}
\end{equation}
where $\pi_{\rm S}={\rm AU}/\ds$.  However, in none of the observed events were all 
three lensing observables measured in full.  As a result, we inferred the physical 
parameters by performing a Bayesian analysis.  This approach incorporates not only 
the constraints from the measured observables but also prior information based on 
the physical and dynamical distributions of lenses, as well as their mass function.

% Figure 9 ------------------------------------------------------
\begin{figure}[t]
\includegraphics[width=\columnwidth]{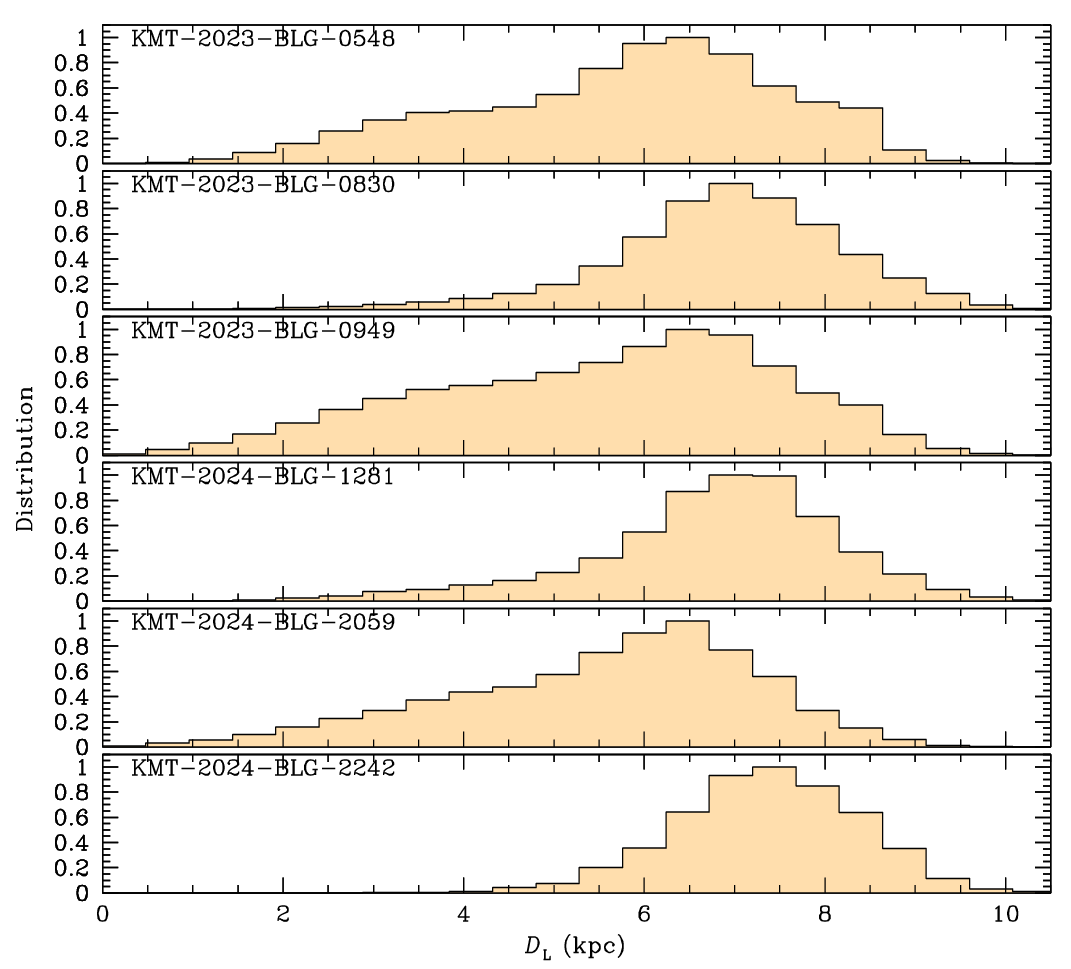}
\caption{
Posteriors for the distance to the lens systems. 
}
\label{fig:nine}
\end{figure}
% --------------------------------------------------------------

The Bayesian analysis commenced with the generation of a large set of synthetic 
lensing events.  The physical parameters of each synthetic event, $(M_i, D_{{\rm L}, i}, 
D_{{\rm S}, i}, \mu_i)$, were assigned using a Monte Carlo simulation based on a prior 
Galactic model and lens mass function.  For the Galactic model, which describes the 
physical and dynamical distributions of lensing objects in the Galaxy, we adopted the 
model outlined in \citet{Jung2021}.  In this model, the Galaxy is basically represented 
as a triaxial bulge plus a double‑exponential disk with some refinement based on Gaia 
data and observed disk rotation curve.  For the lens mass function, which characterizes 
the distributions of lens masses, we employed the model detailed in \citet{Jung2022}, 
which includes contribution from stars, brown dwarfs, and remnants (white dwarfs, neutron 
stars, and black holes).  For each synthetic event, we calculated the lensing observables 
corresponding to the assigned physical parameters using the relations in Eqs.~(\ref{eq4}).  
Based on the ensemble of generated synthetic events, the posterior probability distributions 
of the lens mass $M$ and distance $\dl$ were derived by weighting each event according to 
its consistency with the observed lensing observables. The applied weight accounts for how 
well the simulated observables match the measured values and is defined as
\begin{equation}
w_i=\exp\left(-{\chi_i^2\over 2}\right); \qquad 
\chi_i^2 = {(t_{{\rm E},i}-\te)^2\over \sigma(\te)^2} + 
{(\theta_{{\rm E},i}-\thetae)^2\over \sigma(\thetae)^2}.  
\label{eq6}
\end{equation}
Here, $(\te, \thetae)$ are the observed Einstein timescale and angular Einstein radius, 
and $\sigma(\te), \sigma(\thetae)$ denote their respective uncertainties.
For events 
without a measured $\thetae$, the second term in the $\chi^2$ calculation in 
Eq.~(\ref{eq6}) is not included.

Figures~\ref{fig:eight} and \ref{fig:nine} present the posterior distributions of the 
lens mass and distance obtained from the Bayesian analysis.  The estimated physical 
parameters, including $M_{\rm h}$, $M_{\rm p}$, $\dl$, and $a_\perp$, are summarized 
in Table~\ref{table:nine}.  Here, $M_{\rm h}$ and $M_{\rm p}$ represent the masses of 
the host star and planet, respectively, while $a_\perp$ denotes the projected separation 
between the planet and its host.  The median value of each parameter was adopted as the 
representative estimate, with the lower and upper limits defined by the 16th and 84th 
percentiles of the posterior distribution, respectively.  The tables also provide the 
probabilities of the lens residing in the Galactic disk ($p_{\rm disk}$) or bulge 
($p_{\rm bulge}$).

For KMT-2023-BLG-0548, the posterior distribution of the lens mass reveals two distinct 
peaks: one at $M_1 \sim 0.12~M_\odot$ (low-mass peak) and another at $M_1 \sim 0.72~M_\odot$ 
(high-mass peak). To account for this bimodal distribution, we separately estimated the 
physical parameters associated with each peak.  Based on the posterior distribution near 
the low-mass peak, the planetary companion is estimated to have a sub-Jovian mass of 
approximately $0.6~M_{\rm J}$. In contrast, the posterior distribution near the high-mass 
peak suggests a significantly more massive planet, with a super-Jovian mass of approximately 
$3.5~M_{\rm J}$. The host star associated with the low-mass peak is an M dwarf located in the 
Galactic disk, whereas the host corresponding to the high-mass peak is a K-type dwarf situated 
in the Galactic bulge.

For the event KMT-2023-BLG-0830, the lens system consists of a planet with a mass of
approximately $\sim 0.06~M_{\rm J}$, comparable to the mass of Neptune in the solar system,
and a host star with a mass of approximately $\sim 0.22~M_\odot$, corresponding to a mid-M
dwarf. The estimated distance to the system, $\dl \sim 7.3$~kpc, strongly suggests that this
planetary system is likely located in the Galactic bulge.

The companion of KMT-2023-BLG-0949L is a giant planet with an estimated mass of about 
$\sim 6~M_{\rm J}$ for the close solution and $\sim 4~M_{\rm J}$ for the wide solution. 
The host star, which has roughly half the mass of the Sun, is situated at a distance of 
approximately $\sim 6$ kpc. The likelihood of this planetary system being located in the 
Galactic disk or bulge is nearly equal.

For KMT-2024-BLG-1281, the lens parameters were estimated based on the intermediate 
solution, which provides a better fit than the alternative models. The planet has a 
mass slightly greater than that of Neptune in the Solar System and orbits an M-dwarf 
host star with a mass of approximately $\sim 0.3~M_\odot$. This planetary system is 
more likely located in the Galactic bulge than in the disk.

The planet KMT-2024-BLG-2059Lb has an estimated mass of about seven times that of Earth, 
placing it in the category of super-Earths. Its host star has a mass of approximately 
$M_{\rm h} \sim 0.5~M_\odot$, consistent with an early M-type dwarf. Based on the 
inferred lens parameters and Galactic model, the probabilities of the planetary system 
being located in the Galactic disk or bulge are roughly equal.

The short timescale and small angular Einstein radius of KMT-2024-BLG-2242 suggest that 
the primary lens has a low mass, estimated at $M_{\rm h} \sim 0.07~M_\odot$. This mass 
falls below the threshold for hydrogen burning, indicating that the host is likely a 
brown dwarf. The planet, with an estimated mass of 17 or 14 Earth masses depending on 
the solution, is similar in mass to Uranus and Neptune in the Solar System. Including 
KMT-2024-BLG-2242Lb, a total of 16 planets orbiting brown dwarf hosts have been discovered 
through microlensing (see Table 3 of \citealt{Han2024c}). This finding underscores the 
capability of microlensing to detect planetary systems around faint or non-luminous 
hosts.

\section{Summary and conclusion} \label{sec:seven}

We carried out detailed analyses of six microlensing events: KMT-2023-BLG-0548, 
KMT-2023-BLG-0830, KMT-2023-BLG-0949, KMT-2024-BLG-1281, KMT-2024-BLG-2059, and 
KMT-2024-BLG-2242. These events were identified in the KMTNet survey data collected 
during the 2023 and 2024 observing seasons, as part of a targeted search for 
microlensing events exhibiting anomalies with durations shorter than one day. 
Such short-lived anomalies may indicate the presence of a planetary companion to 
the lens, prompting a detailed investigation to determine the nature and origin 
of the observed perturbations.

We modeled the observed light curves using various lens-system configurations 
that could produce short-term anomalies, including a planetary companion to the 
lens, a very wide or close binary lens companion, and a faint source companion. 
We also considered the potential for multiple interpretations due to lensing 
degeneracies. We found that planetary companions to the lenses are responsible 
for the anomalies in all analyzed events. In the case of the two events 
KMT-2024-BLG-2059 and KMT-2024-BLG-2242, the duration was short due to the small 
caustic size induced by low-mass planet. However, the short durations of anomalies 
in other events arose from different causes. In the case of KMT-2023-BLG-0548, 
the anomaly was brief due to the substantial deviation of the planetary separation 
from the Einstein radius, leading to a small caustic size. For KMT-2023-BLG-0830 
and KMT-2024-BLG-1281, the short duration of the anomaly was caused by a combination 
of a small planet-to-host mass ratio and the restricted crossing area of the source, 
resulting from the elongated shape of the caustic. In KMT-2023-BLG-0949, the source 
traversed only a small portion of the caustic, which resulted in the short-lived 
anomaly.

We determined the physical parameters of the lens systems through Bayesian analysis, 
combining constraints from the observed lensing parameters with priors based on a 
Galactic model and the lens mass function to estimate the most probable values. For 
KMT-2023-BLG-0548, the Bayesian posterior shows two distinct mass peaks: the lower-mass 
peak suggests a sub-Jovian planet orbiting an M dwarf in the Galactic disk, while the 
higher-mass peak indicates a super-Jovian planet around a K-type dwarf in the bulge. 
For KMT-2023-BLG-0830, the lens system comprises a Neptune-mass planet orbiting an M 
dwarf, located in the Galactic bulge. For KMT-2023-BLG-0949, the lens system consists 
of a super-Jovian planet orbiting a $\sim0.5~M_\odot$ host at a distance of roughly 6 
kpc. For KMT-2024-BLG-1281, the lens is consisted of a planet with a mass slightly 
greater than that of Neptune and and an M-dwarf host star. The planet KMT-2024-BLG-2059Lb, 
with a mass about seven times that of Earth, is classified as a super-Earth and orbits 
an M dwarf of $\sim0.5~M_\odot$. The short timescale and small angular Einstein radius 
of KMT-2024-BLG-2242 suggest a $\sim 0.07~M_\odot$ host, likely a brown dwarf, with a 
planet comparable in mass to Uranus or Neptune.

Although the physical parameters of the lens in the analyzed microlensing events are 
subject to considerable uncertainty due to incomplete measurements of the lensing 
observables, these uncertainties can be significantly mitigated through future 
high-resolution follow-up observations.  Instruments such as the Keck Adaptive Optics 
(AO) system on the 10-meter Keck Telescope and the upcoming 30-meter European Extremely 
Large Telescope (E-ELT), which is expected to become operational around 2030, will 
facilitate such observations. High-resolution imaging will enable the spatial resolution 
of the lens and source, allowing for direct confirmation of microlensing models by 
comparing the lens-source relative proper motion inferred from light curve modeling 
with that measured from follow-up data. Furthermore, resolving the lens will provide 
tighter constraints on the properties of the planetary host and help to break degeneracies 
in the estimates of lens mass and distance, particularly in cases with multiple degenerate 
solutions, such as KMT-2023-BLG-0548.

Among the six newly discovered planets, four are found to have masses significantly lower 
than that of Jupiter in our solar system. Recently, \citet{Zang2025} conducted a statistical 
analysis of planet-to-host mass ratios using a sample of exoplanets discovered through the 
KMTNet survey, revealing that planets in Jupiter-like orbits exhibit a bimodal distribution, 
with distinct peaks corresponding to super-Earths and gas giants. Of the low-mass planets 
reported in this study, KMT-2024-BLG-2059Lb falls into the super-Earth category, while the 
other three (KMT-2023-BLG-083Lb, KMT-2024-BLG-1281Lb, and KMT-2024-BLG-2242Lb) are comparable 
in mass to Uranus-like planets in our solar system. Although the Zang et al. study was based 
on a total of 63 planets, this sample size remains relatively limited. The newly reported 
planets are therefore expected to contribute meaningfully to refining the statistical analysis 
and improving the robustness of conclusions about the demographics of cold exoplanets.

% --------------------------------------------------------------
\begin{acknowledgements}
% KMTNet
This research has made use of the KMTNet system operated by the Korea Astronomy and Space Science 
Institute (KASI) at three host sites of CTIO in Chile, SAAO in South Africa, and SSO in Australia. 
Data transfer from the host site to KASI was supported by the Korea Research Environment Open NETwork 
(KREONET). 
% KASI support for Han 
C.Han acknowledge the support from the Korea Astronomy and Space Science Institute under the R\&D program 
(Project No. 2025-1-830-05) supervised by the Ministry of Science and ICT.
% Yee
J.C.Y., I.G.S., and S.J.C. acknowledge support from NSF Grant No. AST-2108414. 
%Yossi Shvartzvald
% Chinese researchers
W.Zang acknowledges the support from the Harvard-Smithsonian Center for Astrophysics 
through the CfA Fellowship.
% MOA
The MOA project is supported by JSPS KAKENHI Grant Number 
JP16H06287, JP22H00153 and 23KK0060.
% Clement Ranc
C.R. was supported by the Research fellowship of the Alexander von Humboldt Foundation.
\end{acknowledgements}

\end{document}